\documentclass[aps,twocolumn,amsfonts,showpacs,pra,amsmath,amssymb,superscriptaddress,longbibliography]{revtex4-2}
\usepackage{graphicx, color}
\usepackage{hyperref}
\hypersetup{colorlinks=true,linkcolor=blue,citecolor=blue,urlcolor=blue}
\usepackage{dcolumn}
\usepackage{bm}
\usepackage{mathtools}
\usepackage{newtxtext}
\usepackage{newtxmath}
\usepackage[version=3]{mhchem}
\usepackage[normalem]{ulem}
\usepackage{lipsum}
\usepackage{braket}
\usepackage{comment}
\usepackage{physics}
\usepackage{anyfontsize}
\usepackage[T1]{fontenc}
\usepackage[utf8]{inputenc}

\DeclareRobustCommand{\Erase}{\bgroup\markoverwith{\textcolor{red}{\rule[.5ex]{2pt}{0.4pt}}}\ULon}

\begin{document}

\title{Theoretical analysis of photon detection mechanism in superconducting single-photon detectors}

\newcommand{\TohokuUniv}{Department of Applied Physics, Tohoku University, Sendai 980-8579, Japan}
\newcommand{\CSIS}{Center for Science and Innovation in Spintronics, Tohoku University, Sendai 980-8577, Japan}

\author{Yusuke Masaki}
\email{yusuke.masaki.c1@tohoku.ac.jp}
\affiliation{\TohokuUniv}

\author{Hiroaki Matsueda}
\email{hiroaki.matsueda.c8@tohoku.ac.jp}
\affiliation{\TohokuUniv}
\affiliation{\CSIS}

\date{\today}

\begin{abstract}
To elucidate the photon detection mechanism of superconducting single-photon detectors, we theoretically examine the dynamics of type-II superconductors with a bias current using the two-dimensional time-dependent Ginzburg-Landau and the Maxwell equations. The photon injection that weakens the superconducting order parameter is treated phenomenologically as a local temperature increase, and the amount of injection is controlled by the initial hotspot radius. The photon is detected by the voltage change between two electrodes attached to the left and right edges of the superconductor. We find that certain parameter ranges can be explained by the traditionally considered hotspot model, while other parameter ranges are governed by the generation and annihilation of superconducting vortex and antivortex pairs. The photon detection is possible for an initial hotspot radius that exceeds a threshold value. We find that the generation of a vortex--antivortex pair occurs near the threshold. The flow of the pair perpendicular to the current direction finally creates a normal region for the photon detection. The voltage change for the Ginzburg--Landau parameter close to the transition point from type-II to type-I superconductor shows anomalous behavior that is not associated with the dynamics of the vortex--antivortex pair. We also examine the effects of spatially non-uniform current density on the voltage change and the superconducting order parameter to provide a hint to understand the behavior of wide-strip single-photon detectors. The estimated values of incident photon energy and response time for photon detection are reasonable in comparison with experiments. The present comprehensive examination provides useful guidelines for flexible design of device structures.
\end{abstract}

\maketitle

\section{Introduction}

The superconducting single-photon detector (SSPD) has many advantages, such as fast response, detection efficiency, low jitter, and low dark count~\cite{Goltsman,Goltsman2,Yamashita,Hadfield,Miki,Miki2,Yamashita2,Yamashita3,Yamashita4,Miki3,Yamashita5}. The extremely high photon detection capability is essential for photonic quantum information technologies. Nevertheless, the photon detection mechanism of the SSPD is not adequately understood. Therefore, currently there are no specific guidelines for the selection of superconducting materials and the design of devices to further improve performance. Typically, SSPD devices are composed of a superconducting nanostrip with a meander pattern (superconducting nanowire single-photon detector, SNSPD), and there are still restrictions regarding the polarization dependence and productivity. In this paper, our aim is to investigate the SSPD mechanism to overcome this situation.

The SNSPD mechanism has been widely believed to be described by a hotspot model. According to this model, the device size for high-performance detection is severely limited, and for this reason the nanostrip structure is usually fabricated. However, recent experiments have shown that single-photon detection can also be performed with a micrometer-sized device~\cite{Korneeva,Yabuno} (superconducting wide-strip photon detector, SWSPD). In particular, the authors in Ref.~\cite{Yabuno} propose a high critical current bank structure to suppress the intrinsic dark count caused by the nonuniform distribution of superconducting currents in the microstrip. The experimental successes for efficient photon detection by using the wide strip provide a good opportunity to reconsider the traditional hotspot model for the SSPD and SNSPD mechanisms.

In the hotspot model, the superconductivity is locally suppressed by the incident photon whose energy scale is two or three orders of magnitude larger than the superconducting gap. In this case, without going into the details of the microscopic electronic states, it can be expected that the suppression is a thermal process. We refer to this process as initial hotspot formation. The superconducting current in the SSPD device flows outside the hotspot region. If the device is composed of a thin wire, the increase in the superconducting current density outside the hotspot region may exceed a critical value, leading to local breaking of the superconducting state. This process can be observed by the voltage between the two wire terminals. Although the formation of the initial hotspot region is plausible, the description of the subsequent processes seems to be insufficient to understand the detection mechanism in the micrometer-sized SSPD devices. In the subsequent processes, the hotspot model focuses on the spatiotemporal density change of quasiparticles or the expansion of the region. There are previous studies on the hotspot dynamics using the kinetic-equation approach or the heat-diffusion equation~\cite{Kadin3,Semenov,Semenov2,Engel,Kozorezov,Marsili,Vodolazov}. However, analyzing the thermal distribution alone is not sufficient for a detailed understanding of electromagnetic response.

An alternative possible mechanism is based on the formation of the vortex--antivortex pair. This idea was originally proposed in Refs.~\cite{Kadin,Kadin2}, but to the best of our knowledge, no detailed study that directly visualizes their dynamics has been reported so far. 
In this vortex--antivortex model, the dynamics of vortices and antivortices perpendicular to the current flow direction result in a voltage change. Vortex-assisted mechanisms were also proposed in Refs.~\cite{Qiu,Bartolf,Hofherr,Bulaevskii2,Bulaevskii,Zotova2,Vodolazov2,Engel2,Saman}, inspired by the single-vortex-crossing mechanism for dark count~\cite{Hofherr,Bulaevskii2}. A recent study~\cite{He} reports a unified theory of dark‐count rate and system‐detection efficiency based on a vortex-crossing framework. The photon detection mechanism based on this model is expected to be less sensitive to the finite-size limitations, consistent with the realization of SWSPD. The creation of vortex--antivortex pairs is also observed by the local heating effect of a scanning tunneling microscope tip~\cite{Ge}. Theoretical confirmation of pair generation would represent an important step toward developing a comprehensive understanding of the physics of SSPD devices. 
At the same time, we aim to clarify how the model parameters differ between the hotspot and vortex--antivortex models.

Here, we summarize two previous theoretical works that are based on the time-dependent Ginzburg--Landau (TDGL) equation and are closely related to the present work.

The hotspot mechanism is discussed in Ref.~\cite{Ota}. The authors treat the three-dimensional TDGL equation coupled with the Maxwell and heat-diffusion equations and take the bias current 95$\%$ of the critical current. The heat diffusion is introduced by taking into account the effect of a heat sink on relaxation. They found a dynamical transition to a resistive state when the incident photon has higher energy than the superconducting transition temperature. 
The authors presented the numerical results for the surface temperature of the initial hotspot close to $T_c$. After the photon injection, a weak-superconducting strip begins expanding. This strip reaches both sample edges, and a tiny normal region appears at the center of the system. The superconductivity in the regions sandwiched between the sample edges and the tiny normal region easily breaks, and finally the system becomes the resistive state.
Unfortunately, the threshold energy for the transition is two to three orders of magnitude larger than the realistic scale. One of the drawbacks would be to take a small GL parameter, $\kappa=\lambda_0/\xi_0$, where $\lambda_0$ and $\xi_0$ are penetration depth and coherence length at zero temperature, respectively. Here, the vortex--antivortex dynamics is not clearly presented, although the authors mention that the vortex--antivortex pairs enter from both the normal region and the edges for sufficiently large incident photon energy. 

The examination of the vortex--antivortex mechanism in terms of the TDGL equation is reported in Ref.~\cite{Zotova}. The authors numerically solved the two-dimensional TDGL equation coupled with the Poisson equation for the electric potential and the heat-diffusion equation. They do not take into account the vector potential for the magnetic field. This assumption is valid in the large-$\kappa$ limit. They take the coherence length at zero temperature, the initial radius of the hotspot, and the width of the film (perpendicular to the current-flow direction) as $\xi_0=7.5$ nm, $R_{\mathrm{init}}=9$ nm, and $w=13\xi_0\sim 52\xi_0$, respectively. The local temperature increase is determined by energy conservation $E_{\mathrm{photon}}=\pi R_{\mathrm{init}}^{2}dC_{v}\Delta T$, where $C_{v}$ is a heat capacity, $d$ is a thickness of the sample, and $\Delta T$ is a temperature increase due to photon injection. At current larger than a threshold (the authors call this detecting current), they found that the superconducting state collapses starting from the appearance of a vortex--antivortex pair in the center of the initial hotspot region. Lorentz force causes their motion that heats the system locally and gives rise to a normal domain. The authors particularly focus on the dependence of detecting current on $\Delta T$ and $w$. In this study, $\xi_0$ is comparable to $R_{\mathrm{init}}$, but we consider that the initial hotspot size strongly affects the stability of the formation of a vortex--antivortex pair. Therefore, the parameters should be determined more carefully. Furthermore, the authors do not comment on why the traditional hotspot mechanism is not consistent with their analysis.

Motivated by the present status mentioned above, we examine the dynamics of type-II superconductors with a bias current using the two-dimensional TDGL and the Maxwell equations. The photon injection is treated phenomenologically by the temperature increase inside the initial hotspot region. The important parameter in this study is the initial hotspot radius by which the amount of photon energy is controlled. We find that certain parameter ranges can be explained by the hotspot model, while other parameter ranges are governed by the generation and annihilation of a vortex--antivortex pair. The photon detection is possible for an initial hotspot radius that exceeds a threshold value. We demonstrate that the pair generation occurs near the threshold. The flow of the pair perpendicular to the current direction finally leads to the formation of a normal region responsible for photon detection. During the formation process, the voltage change shows oscillation that originates from repeated penetration of vortices and antivortices from the sample edges. The voltage change for the GL parameter close to the transition point from type-II to type-I superconductor shows anomalous oscillations that are not associated with the vortex--antivortex pair. We also examine the effects of spatially non-uniform current density on the voltage change and the superconducting order parameter in order to understand the behavior of wide-strip single-photon detectors. The estimated values of incident photon energy and response time for photon detection are reasonable in comparison with experiments. The present comprehensive examination provides useful guidelines for flexible design of device structures.

The organization of this paper is as follows. In Sect.~\ref{model}, we introduce our model and the calculation method. In Sect.~\ref{results}, we show our numerical results. Finally, we summarize our conclusion in Sect.~\ref{conclusion}.

\section{Model}
\label{model}

\begin{figure}[tb]
\begin{center}
\includegraphics[width=\hsize]{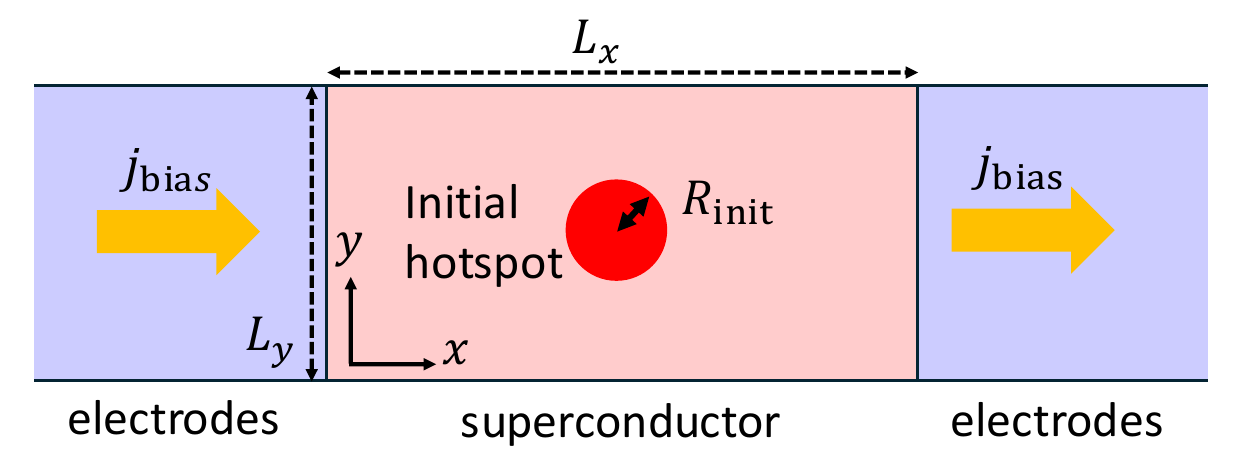}
\end{center}
\caption{
Illustration of our system. The red and blue areas correspond to superconductor and metallic electrodes, respectively. The system size and current-flow direction are indicated, respectively. The superconducting region is given by $0\le x\le L_{x}$ and $0\le y\le L_{y}$. The center of the initial hotspot with radius $R_{\mathrm{init}}$ is set to the center of the superconductor.
}
\label{SSPDfig1}
\end{figure}

Figure~\ref{SSPDfig1} illustrates our system. The system is composed of a two-dimensional superconductor at zero temperature ($T_{\mathrm{sys}}=0$) with a size of $L_{x}\times L_{y}=40\xi_0\times 20\xi_0$, where $\xi_0$ is the coherent length of the superconductor at zero temperature. In ultrathin film NbN with the film size of $4\sim 5$ nm, $\xi_0$ is about $5$ nm, and thus the width of the present system is about $100$ nm. The left and right terminals are connected to metallic electrodes and the upper and lower edges are connected to vacuum. The bias current $j_{\mathrm{bias}}$ is injected from the left electrode. In most of the numerical simulations, we take $\kappa=10\gg 1/\sqrt{2}$, and thus the system is in the type-II superconductor.

In order to study the dynamics of the order parameter and electromagnetic fields, we solve the TDGL and the Maxwell equations, which are given, respectively, by
\begin{align}
&-\frac{\hbar^{2}}{2m^{\ast}D}\left(\frac{\partial}{\partial t}-i\frac{e^{\ast}}{\hbar}\varphi\right)\Delta \nonumber \\
&\;\; = -\alpha\left(T\left(\vb*{r}\right)\right)\Delta+\beta\left|\Delta\right|^{2}\Delta+\frac{1}{2m^{\ast}}\left(-i\hbar\vb*{\nabla}+\frac{e^{\ast}}{c}\vb*{A}\right)^{2}\Delta, \label{eq:tdgl}  \\
&\vb*{\nabla}\times\left( \vb*{\nabla} \times \vb*{A} \right) = \frac{4\pi}{c}\vb*{j}, \label{eq:Maxwell}
\end{align}
with the current density $\vb*{j}$ defined by
\begin{align}
&\vb*{j}=-\frac{e^{\ast}}{m^{\ast}}\Re\left[\Delta^{\ast}\left(-i\hbar\vb*{\nabla}+\frac{e^{\ast}}{c}\vb*{A}\right)\Delta\right]+\sigma\left(-\vb*{\nabla}\varphi-\frac{1}{c}\frac{\partial\vb*{A}}{\partial t}\right), \label{eq:current}
\end{align}
where the first and second terms of $\vb*{j}$ correspond to superconducting and normal currents, denoted by $\vb*{j}_{s}$ and $\vb*{j}_{n}=\sigma\vb*{E}$, respectively. Here, $\Delta$, $\varphi$, $\vb*{A}$, $D$, $\sigma$, $e^{\ast} (>0)$, and $m^{\ast}$ are the superconducting order parameter, scalar potential, vector potential, diffusion constant, electric conductivity in the normal region, and effective mass and effective charge of a superconducting electron, respectively. The parameter $\alpha$ is determined phenomenologically by $\alpha\left(T\left(\vb*{r}\right)\right)=\alpha(0)\left(1-T(\vb*{r})/T_{c}\right)$, where $T(\vb*{r})$ is the space-dependent temperature due to the formation of the initial hotspot region and subsequent dynamics, $T_{c}$ is the critical temperature and $\alpha_{0}$ is a positive constant. The parameter $\beta$ is a positive constant. By a phase redefinition of the order parameter and the gauge transformation of the vector potential $\vb*{A}$, we will eliminate the scaler potential $\varphi$.

We summarize the boundary conditions employed in our numerical simulations. The upper and lower edges of the superconducting sample are connected to vacuum, and the left and right edges of the superconducting sample are connected to metallic electrodes. The Neumann boundary condition is imposed at superconductor/vacuum interfaces:
\begin{align}
\vb*{e}_{y}\cdot\left.\left(-i \hbar\vb*{\nabla}+\frac{e^{\ast}}{c}\vb*{A}\right) \Delta\right|_{y=0,L_{y}}=0,
\end{align}
where $\vb*{e}_{\mu=x,y,z}$ is a unit vector along the $\mu$ direction. At the left and right edges of the superconductor, we consider the following condition for the order parameter:
\begin{align}
\left.\Delta\right|_{x=0,L_{x}}=0.
\end{align}
We assume that the magnetic field, induced by the transport current, is along the $z$ direction, $\bm{B} = B_z \bm{e}_z = \bm{\nabla}\times \bm{A}$, and consider the following condition for $B_z$. At the upper ($+$) and lower ($-$)  superconductor/vacuum interfaces, the magnetic fields are given by
\begin{align}
\left.B_{z}\right|_{y=0,L_{y}}=\pm\frac{L_{y}}{2}\frac{4\pi}{c}j_{\mathrm{bias}},
\end{align}
where the bias current $j_{\mathrm{bias}}$ satisfies the following conditions:
\begin{align}
j_{\mathrm{bias}}&=\frac{1}{L_{y}}\int_{0}^{L_{y}}j_{x}(x=0,y)dy \nonumber \\
&=\frac{1}{L_{y}}\int_{0}^{L_{y}}\left(j_{n}\right)_{x}(x=0,y)dy.
\end{align}
We assume a spatially uniform normal current at both the metallic electrodes [i.e., $j(x=0, y)=j(x=L_x,y) = j_{\mathrm{bias}}$], and the boundary condition on the magnetic fields there is given by
\begin{align}
\left. B_z\right|_{x=0,L_{x}}=\frac{4\pi}{c}j_\mathrm{bias}y. \label{eq:B}
\end{align}

In this study, the photon injection is treated phenomenologically. A simple method is to consider heating in a fixed area due to the photon injection. We introduce the following temperature difference between the inside and outside of the initial hotspot:
\begin{align}
T(\vb*{r}) = 
\begin{cases}
    T_{\mathrm{sys}}+\Delta T  & 0 \le |\vb*{r} -\vb*{r}_c| \le R_{\mathrm{init}} \\
    T_{\mathrm{sys}} & |\vb*{r} - \vb*{r}_c| > R_{\mathrm{init}} \label{eq:MODEL1_hotspot_temp}
\end{cases},
\end{align}
where $R_{\mathrm{init}}$ is the radius of the circular hotspot initially created by photon injection, $\vb*{r}_c = (L_x/2, L_y/2)$ denotes the center of the initial hotspot, and $\Delta T$ is the temperature increase by photon injection. We take $T_{\mathrm{sys}}=0$ and $\Delta T=T_{c}$. 

Let us define the voltage change $\Delta V$ between the left and right terminals of the superconducting sample by the path integral of the longitudinal component of the electric field as $\Delta V = -\int_{\bm{r}_R}^{\bm{r}_L}\bm{E}_{L}\cdot d\bm{r}$, where $\bm{r}_{L}=\left(3\xi_{0},L_{y}/2\right)$ and $\bm{r}_{R}=\left(L_{x}-3\xi_{0},L_{y}/2\right)$. The measurement positions of $\Delta V$ are taken at distance $3\xi_{0}$ from the terminals to avoid the effects of contact resistance on $\Delta V$. Because $\bm{E}_L$ can be written with a scalar field $V$ as $\bm{E}_{L} = -\bm{\nabla}V$, the integral is independent of the path connecting $\bm{r}_L$ and $\bm{r}_R$, i.e., $\Delta V = V(\bm{r}_L) - V(\bm{r}_R)$. The equation for $V$ is given by
\begin{align}
\nabla^2 V = -\bm{\nabla}\cdot{\bm{E}},
\end{align}
with the boundary conditions
\begin{align}
    \left.\dfrac{\partial V}{\partial x}\right|_{x=0, L_x} = -\dfrac{j_{\mathrm{bias}}}{\sigma},\quad \left.\dfrac{\partial V}{\partial y}\right|_{y=0, L_y} = 0.
\end{align}

Let us introduce the following normalization for convenience of numerical simulation: $\tilde{\vb*{\nabla}}=\vb*{\nabla}/\xi_0$, $\tilde{t}=t/t_{0}$, $\tilde{T}=T/T_{c}$, $\vb*{\tilde{A}}=\vb*{A}/\xi_{0}H_{c2}$, and $\tilde{\Delta}=\Delta/\Delta_{\infty}(0)$, where $t_0 = \xi_0^2/(12D)=4\pi \sigma\lambda_0^2/c^2$, $H_{c2} = c\hbar/e^{\ast}\xi_{0}^{2}$ is the upper critical field, and $\Delta_{\infty}(0) =\sqrt{m^\ast c^{2}/(4\pi e^{\ast 2}\lambda_{0}^{2})}$ is  the zero-temperature order parameter in the absence of the magnetic field. Note that using the parameters in the TDGL equation, $\Delta_{\infty}(0) = \sqrt{\alpha(0)/\beta}$, $\xi_0 =\hbar/\sqrt{2m^\ast \alpha(0)} $, and $\lambda_0 = \sqrt{m^\ast c^2/(4\pi e^{\ast 2})}(\beta/\alpha(0))$. 
A typical value of $t_{0}=\xi_{0}^{2}/12D$ for NbN is about $4.2\times 10^{-2}$ ps (we assume $\xi_0=5.0$ nm and $D=0.5$ cm${}^{2}$/s). Then, the TDGL and the Maxwell equations are normalized as follows:
\begin{align}
&\pdv{\tilde{\Delta}}{\tilde{t}} = -\frac{1}{12}\left[ \left(-i\vb*{\tilde{\nabla}}+\vb*{\tilde{A}}\right)^2\tilde{\Delta}-\left(1-\tilde{T}(\vb*{\tilde{r}})\right)\tilde{\Delta}+|\tilde{\Delta}|^2\tilde{\Delta} \right], \label{eq:tdgl_dimless}  \\
&\kappa^2\vb*{\tilde{\nabla}}\times\left(\vb*{\tilde{\nabla}}\times\vb*{\tilde{A}}\right) = \Re\left[\tilde{\Delta}^{*} \left(-i\vb*{\tilde{\nabla}}+\vb*{\tilde{A}}\right) \tilde{\Delta}\right] - \pdv{\vb*{\tilde{A}}}{\tilde{t}}. \label{eq:Maxwell_dimless}
\end{align}

To preserve the gauge invariance of these equations even after discretization, we employ the link-variable method~\cite{Kato}. In the numerical simulation, the system is discretized into $N_{x}\times N_{y}$ meshes with a mesh spacing of $\delta = 0.25\xi_0$ in both directions, i.e., $\delta = L_x/N_x = L_y/N_y$. For even integers $N_x$ and $N_y$, the numerical grid points are defined as $\bm{r}_{i_x,i_y} = (i_x - 1/2, i_y-1/2)\delta$, where $i_{\mu = x,y} = 1,\cdots,N_{\mu}$. Note that there are no grid points located at $x = L_x/2$ or $y = L_y/2$ due to the above discretization; in the following calculations, these positions  are given by $(L_x - \delta)/2$ and $(L_y - \delta)/2$, respectively; nevertheless, for simplicity, we denote them as $L_x/2$ and $L_y/2$, as long as no confusion arises. In particular, as shown below, such a shift in $\bm{r}_c$ leads to asymmetric dynamics. However, this asymmetry is not essential for photon detection, and we have confirmed that the symmetric dynamics are recovered when the hotspot center is set exactly at the system center, $\bm{r}_c = (L_x/2, L_y/2)$.

\section{Results}
\label{results}

\subsection{Dependence of $R_{\mathrm{init}}$ on voltage change}

Let us present numerical results. We first examine the dependence of $R_{\mathrm{init}}$ on the photon detection performance for $j_{\mathrm{bias}}/j_{0}=0.28$, $j_{0}=(e^{\ast}\hbar/m^{\ast}\xi_0)\Delta_{\infty}^{2}(0)$, and $\kappa=10$. 
The bias current density $j_{\mathrm{bias}}$ is set to be slightly lower than the critical current density $j_c$, where $0.30<j_c/j_0<0.31$. The photon detection performance is characterized by the time evolution of the voltage change $\Delta V$, which is normalized by $V_0 = H_{c2}\xi_{0}^2/t_0$. Here, we do not consider any thermal relaxation processes in which $\Delta V$ again goes to zero toward the next photon detection, since we particularly focus on the detection mechanism. We believe that our conclusions do not change so much even if we additionally introduce these thermal relaxation processes.

\begin{figure}[tb]
\begin{center}
\includegraphics[width=\hsize]{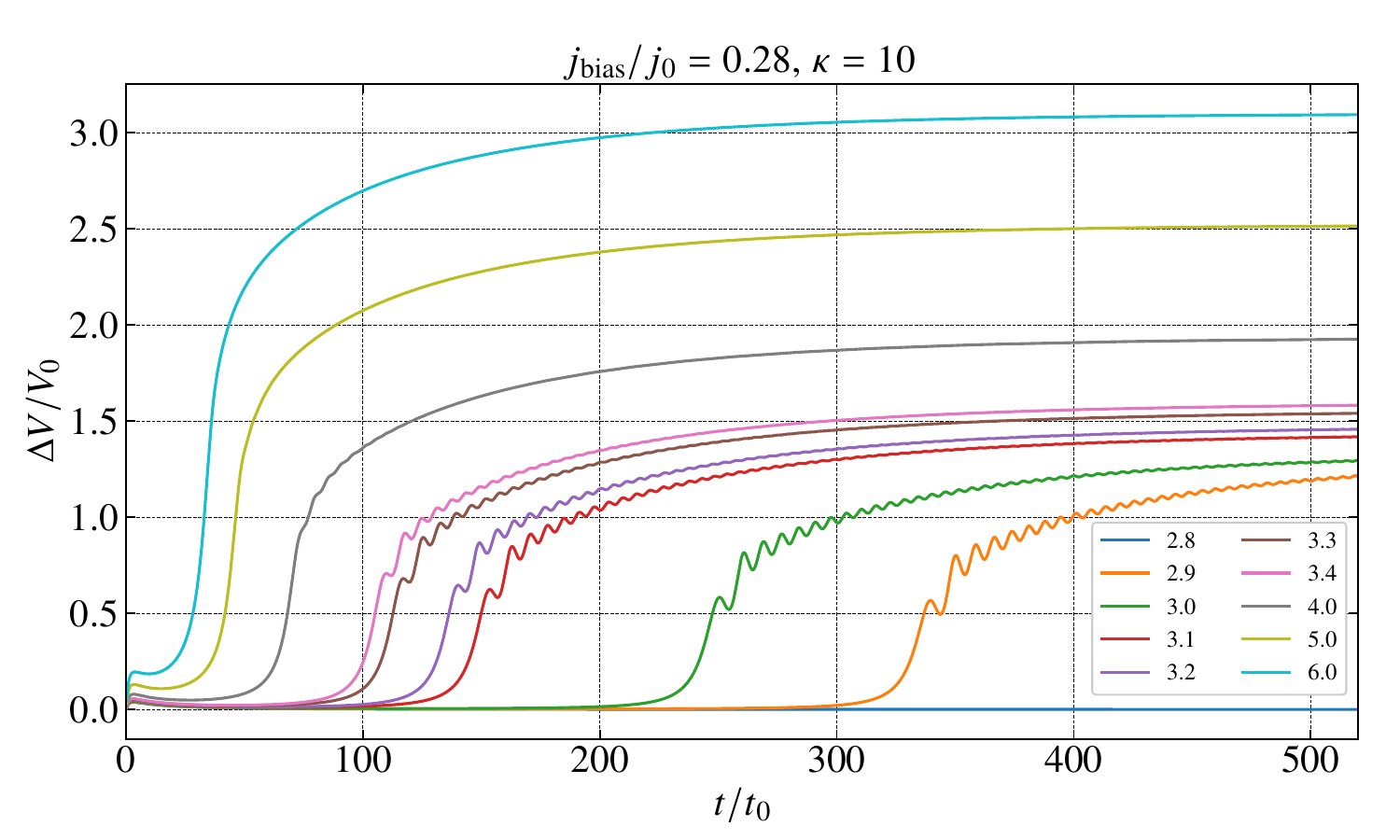}
\end{center}
\caption{
Time evolution of voltage change $\Delta V$. Various lines present the profiles with different values of $R_{\mathrm{init}}/\xi_0$. We take $j_{\mathrm{bias}}/j_{0}=0.28$ and $\kappa=10$.
}
\label{SSPDfig2}
\end{figure}

As shown in Fig.~\ref{SSPDfig2}, for $R_{\mathrm{init}}/\xi_0 \ge 2.9$, $\Delta V$ exhibits a rapid increase at a certain time and subsequently saturates, indicating successful photon detection. We refer to this time as the response time. 
For example, the response time is about $t/t_{0}=250$ for $R_{\mathrm{init}}/\xi_0=3.0$ (see green line in Fig.~\ref{SSPDfig2}). The time for NbN is estimated as $t\sim 10$ ps. The response time depends on $R_{\mathrm{init}}$: a smaller value of $R_{\mathrm{init}}$ leads to a slower response time. For $R_{\mathrm{init}}/\xi_0>3.3$, the response time is smaller than $t/t_{0}=100$. On the other hand, for $R_{\mathrm{init}}/\xi_0<3.3$, the response starts to become noticeably slow. The magnitude of $\Delta V$ after saturation becomes smaller as $R_{\mathrm{init}}$ decreases. In Fig.~\ref{SSPDfig2}, we also observe the oscillatory behavior of $\Delta V$ after the response time. This oscillation becomes more visible for smaller values of $R_{\mathrm{init}}$. We will address the origin of this oscillation in the next subsections.

\begin{figure}[tb]
\begin{center}
\includegraphics[width=0.9\hsize]{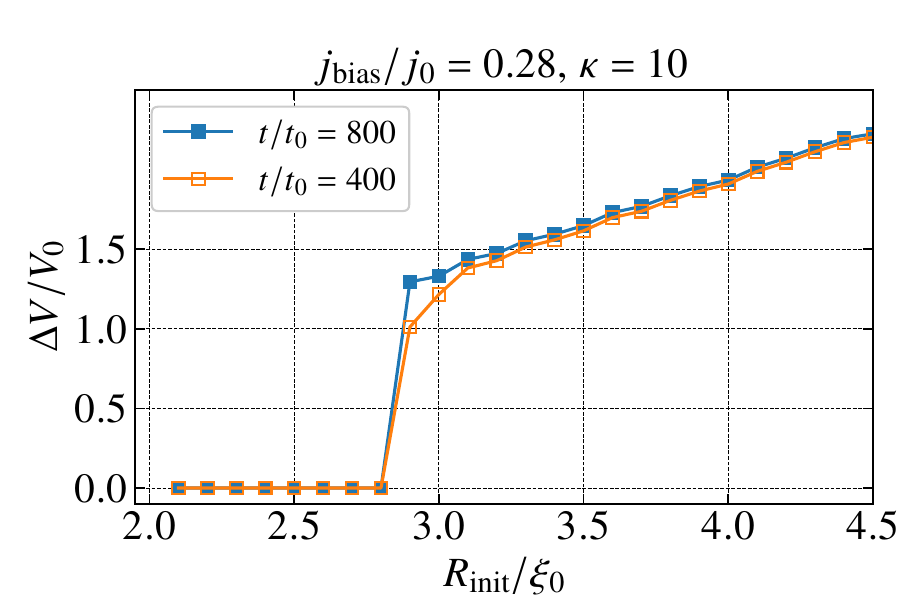}
\end{center}
\caption{
Voltage change $\Delta V$ at $t/t_{0}=400$ (open squares) and $t/t_0 = 800$ (filled squares) as a function of the initial hotspot radius $R_{\mathrm{init}}$. We take $j_{\mathrm{bias}}/j_{0}=0.28$ and $\kappa=10$.
}
\label{SSPDfig3}
\end{figure}

The trends in response time and the magnitude of $\Delta V$ as a function of $R_{\mathrm{init}}$ shown in Fig.~\ref{SSPDfig2} suggest that there exists a threshold value of $R_{\mathrm{init}}$ for photon detection. To see the threshold, we plot $\Delta V$ as a function of $R_{\mathrm{init}}$ in Fig.~\ref{SSPDfig3}. All data are taken at $t/t_{0}=400$ and $t/t_{0}=800$. A sharp threshold is found: $\Delta V$ becomes finite for the parameter range $R_{\mathrm{init}}/\xi_0\ge 2.9$, whereas we have confirmed that, for $R_{\mathrm{init}}/\xi_0 = 2.8$, the resistive state is not attained even up to $t/t_0 = 1600$.

\subsection{Dynamics of amplitude and phase of superconducting order parameter}

\begin{figure}[tb]
\begin{center}
\includegraphics[width=\hsize]{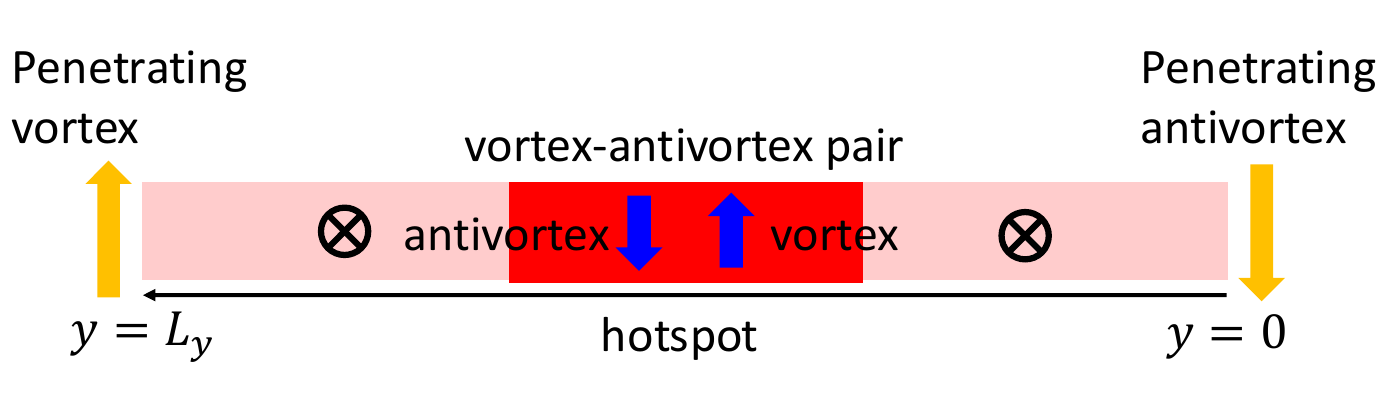}
\end{center}
\caption{
Schematic view of a vortex--antivortex pair generated inside of the hotspot region as well as a penetrating vortex and a penetrating antivortex from outside the superconductor. Cross symbols inside circles indicate the current flow direction.
}
\label{SSPDfig4}
\end{figure}

The oscillation of $\Delta V$ observed in Fig.~\ref{SSPDfig2} originates from the motion of a vortex--antivortex pair generated in the initial hotspot region and from their interactions with vortices and antivortices penetrating the superconductor from outside. Here, 
the latter votices and antivortices are called penetrating votices and penetrating antivortices, respectively, and they are distinguished from the vortex--antivortex pair generated in the hotspot region. Figure~\ref{SSPDfig4} is a schematic viewgraph of these vortices and antivortices.

\begin{figure}[tb]
\begin{center}
\includegraphics[width=\hsize]{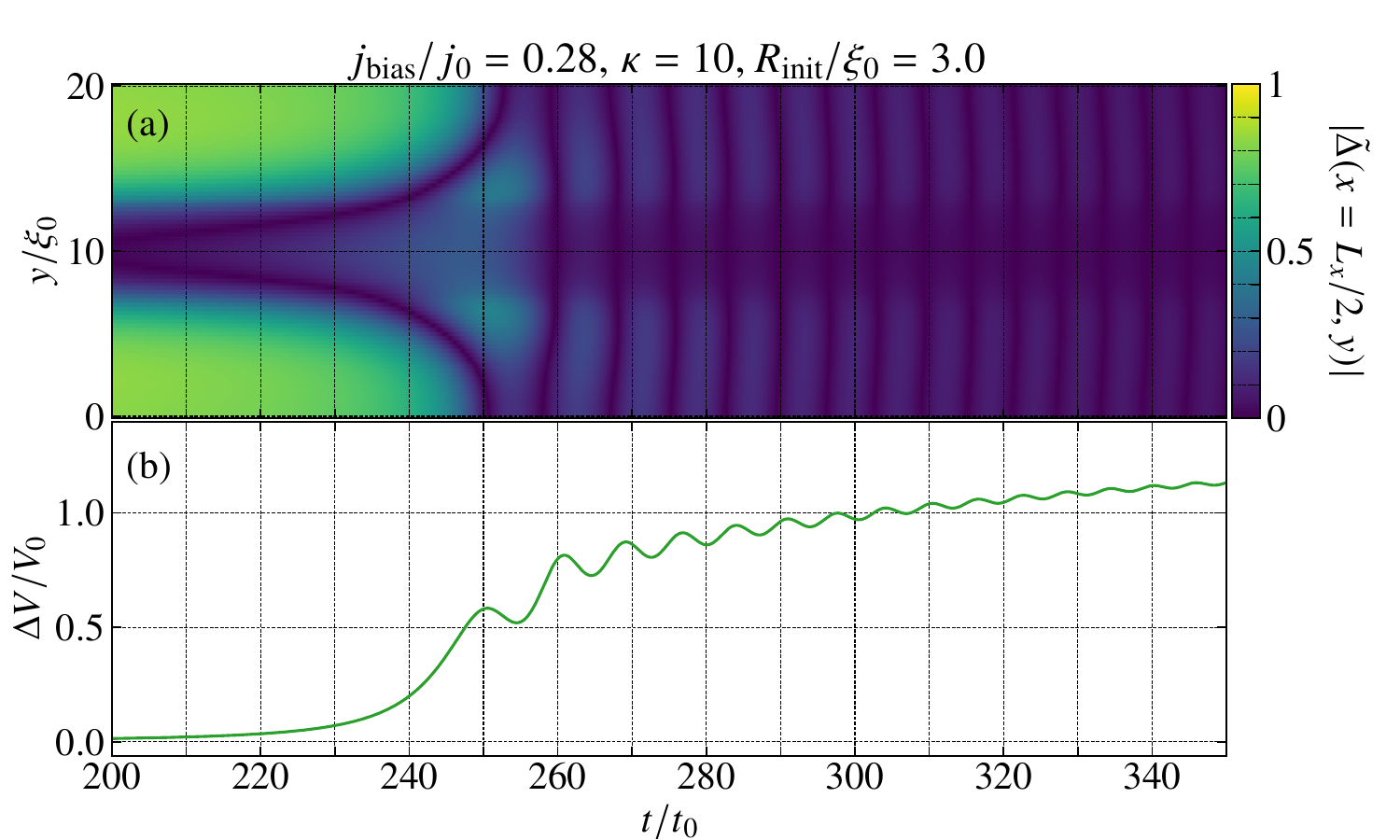}
\end{center}
\caption{
Time evolution of spacetime profile of $\left|\tilde{\Delta}\left(x=L_{x}/2,y\right)\right|$ [(a) upper panel] and that of $\Delta V$ [(b) lower panel] for $R_{\mathrm{init}}/\xi_0=3.0$, $j_{\mathrm{bias}}/j_0=0.28$, $\kappa=10$, and $200\le t/t_{0}\le 350$. Panel (b) is a magnified view of Fig.~\ref{SSPDfig2} for $R_{\mathrm{init}}/\xi_0 = 3.0$. The blight region in panel (a) corresponds to stable superconducting state. The position, $y/\xi_0=10$, corresponds to the hotspot center.
}
\label{SSPDfig5}
\end{figure}

Let us examine the origin of the oscillation in terms of the order parameter. For this purpose, we first present the time evolution of $\left|\tilde{\Delta}\left(x=L_{x}/2,y\right)\right|$ (passing through the hotspot area) in Fig.~\ref{SSPDfig5}(a), where we take $R_{\mathrm{init}}/\xi_0=3.0$, $j_{\mathrm{bias}}/j_0=2.8$, and $\kappa=10$. 
For $t/t_0\le 250$, there are no significant changes outside the initial hotspot region, but a pair of localized dark spots begins to develop inside the hotspot. They are also localized in the $x$ direction. These dark spots propagate along the $y$ axis and reach the upper and lower edges of the superconductor around $t/t_0=250$. After $t/t_0=250$, two dark spots penetrate from the lower and upper edges of the superconductor and propagate toward each other, eventually merging near the center of the system. These processes occur repeatedly. Note that the dynamics is spatially asymmetric about $y=L_y/2$. This is due to the slight shift of the initial hotspot center relative to the system center, introduced by the lattice discretization as mentioned at the end of Sect.~\ref{model}. We have confirmed numerically that this asymmetry does not occur when the hotspot region is located such that its center coincides with the center of the system.

For comparison, we also plot $\Delta V$ in the same time domain as Fig.~\ref{SSPDfig5}(b). Around $t/t_0=250$, $\Delta V$ shows its first local maximum. At the same time, the dark spots generated within the initial hotspot region reach the sample edges, as shown in Fig.~\ref{SSPDfig5}(a). The other local maxima in $\Delta V$ appear, when the pairs of the penetrating dark spots in Fig.~\ref{SSPDfig5}(a) merge with each other near the center of the sample. The penetration of each pair occurs between times when $\Delta V$ reaches its local minimum and its subsequent local maximum. As shown in the next paragraph, the dark spots originates in the vortex--antivortex pair generated in the initial hotspot for $t/t_0 \lesssim 250$ and in the penetrating vortices and antivortices for $t/t_0 \gtrsim 250$.

\begin{figure}
\begin{center}
\includegraphics[width=\hsize]{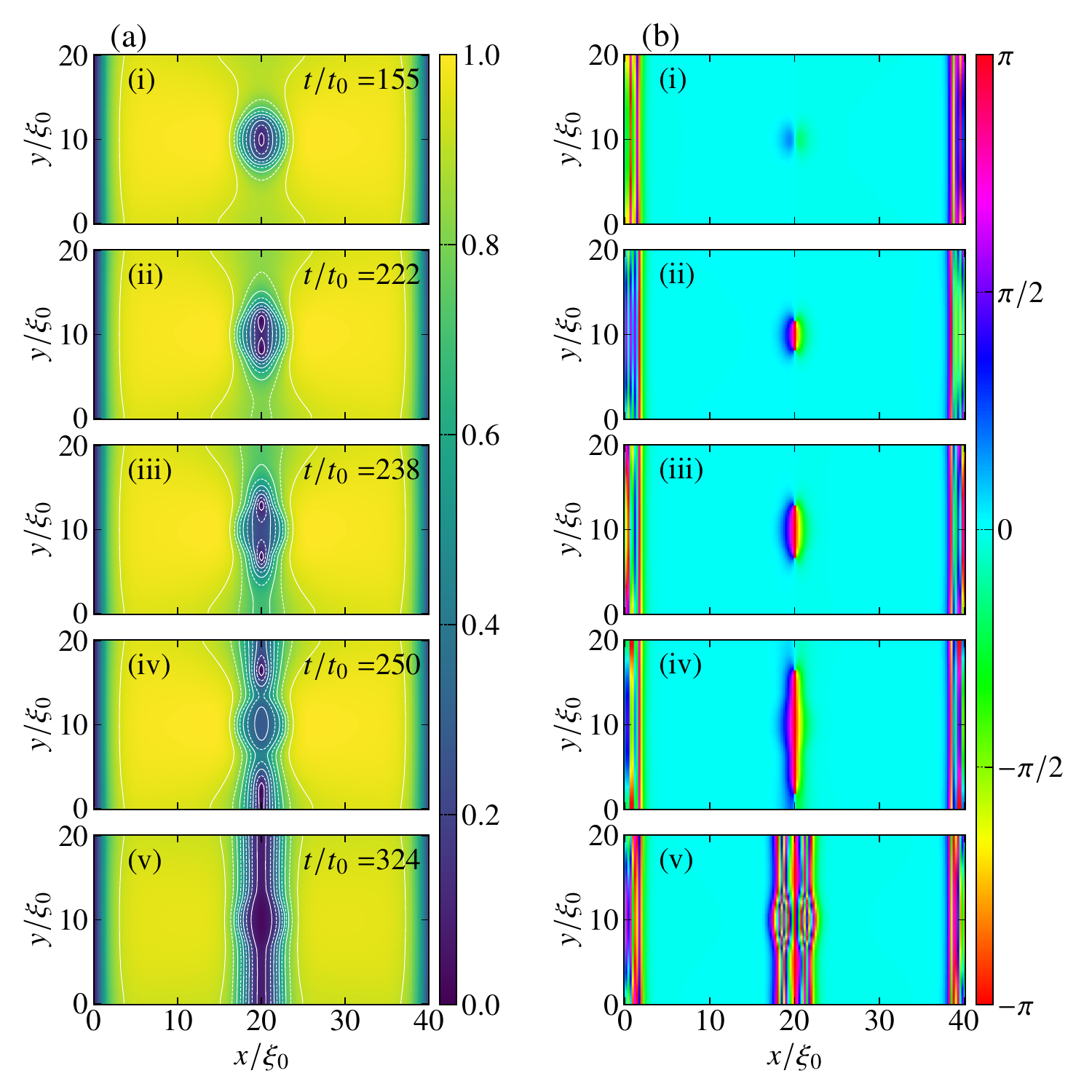}
\end{center}
\caption{
Dynamics of amplitude [(a) left panels] and phase [(b) right panels] of superconducting order parameter. We take $t/t_{0}=155,222,238,250,324$ from top to bottom, $R_{\mathrm{init}}/\xi_0=3.0$, $j_{\mathrm{bias}}/j_0=0.28$, and $\kappa=10$.
In the panels, contours are drawn every 0.1 for $3\xi_0 \le x \le L_x -3\xi_0$; solid (dashed) lines indicate levels of 0.1, 0.3, 0.5, 0.7, and 0.9 (0.2, 0.4, 0.6, and 0.8).
}
\label{SSPDfig6}
\end{figure}

To visualize the generation of the vortex--antivortex pair in the initial hotspot region and the subsequent formation of the normal region, we present in Fig.~\ref{SSPDfig6}(a) color maps of the amplitude of the superconducting order parameter $\tilde{\Delta}(x,y)$ at several representative times, obtained for $R_{\mathrm{init}}/\xi_0=3.0$, $j_{\mathrm{bias}}/j_0=0.28$, and $\kappa=10$. We find that the normal region is finally formed across the upper and lower ends of the sample, and this is consistent with the voltage change in Fig.~\ref{SSPDfig3}. During this process, we observe two small dark spots generated in the initial hotspot and moving separately toward the ends of the sample [see panels (ii)--(iv) of Fig.~\ref{SSPDfig6}(a)]. The radii of these spots are smaller than $R_{\mathrm{init}}/\xi_0=3.0$, and are comparable in size to the vortex core (of order $\xi_0$). After the dark spots reach the sample edges around $t/t_0=250$, the solid normal region is finally formed across the upper and lower edges [see Fig.~\ref{SSPDfig6}(a)(v)]. Our observation cannot be simply understood in terms of the hotspot model.

We also plot the corresponding phases of the order parameter in Fig.~\ref{SSPDfig6}(b). We clearly observe that the two small dark spots correspond to quantum vortices with opposite phase windings ($\pm 2\pi$ phase rotations) [see panels (ii)--(iv) of Fig.~\ref{SSPDfig6}(b)]. Thus, we conclude that they are a vortex--antivortex pair. 
We find that the intermediate region between the two singularities expands as $t$ increases. At $t/t_0=250$ [see Fig.~\ref{SSPDfig6}(b)(iv)], this region is almost in contact with the lower edge of the sample. According to Fig.~\ref{SSPDfig2} or Fig.~\ref{SSPDfig5}(b), a sharp increase in $\Delta V$ occurs and $\Delta V$ takes its first local maximum around this time. We cannot clearly observe the presence of penetrating vortices and antivortices at least up to $t/t_0\le250$. After $t/t_0=250$, $\Delta V$ starts oscillating, as shown in Fig.~\ref{SSPDfig2}.

By careful analysis of the time evolution of the order parameter after $t/t_0=250$ (not shown here), we have confirmed repeated penetration of vortex-antivortex pairs from the outside of the superconductor, and this is the origin of the penetrating dark spots in Fig.~\ref{SSPDfig5}(a). For instance, the first local minimum and the second local maximum in $\Delta V$ are located around $t/t_0=254.5$ and $t/t_0=261$, respectively, and the vortex and antivortex penetrate from the outside around $t/t_0=258$. The time when $\Delta V$ reaches its local minimum corresponds to the onset of the decrease in the edge order parameter, serving as a precursor to the penetration of a vortex or an antivortex, while its local maximum is reached after the penetrating vortex and antivortex annihilate near the center.

\subsection{Parameter region where the hotspot model is plausible}

As mentioned in the previous subsections, the numerical results for smaller $R_{\mathrm{init}}$ values ($R_{\mathrm{init}}/\xi_0\sim 3.0$) and for $\kappa=10$ suggest the vortex--antivortex mechanism for photon detection. In Fig.~\ref{SSPDfig2}, we have already found that the voltage change shows a rather smooth curve for larger $R_{\mathrm{init}}$. The absence of the oscillation may lead to a photon detection mechanism different from the vortex--antivortex pair dynamics. By focusing on the case of larger $R_{\mathrm{init}}$, we demonstrate a hotspot-like behavior for photon detection.

\begin{figure}[tb]
\begin{center}
\includegraphics[width=\hsize]{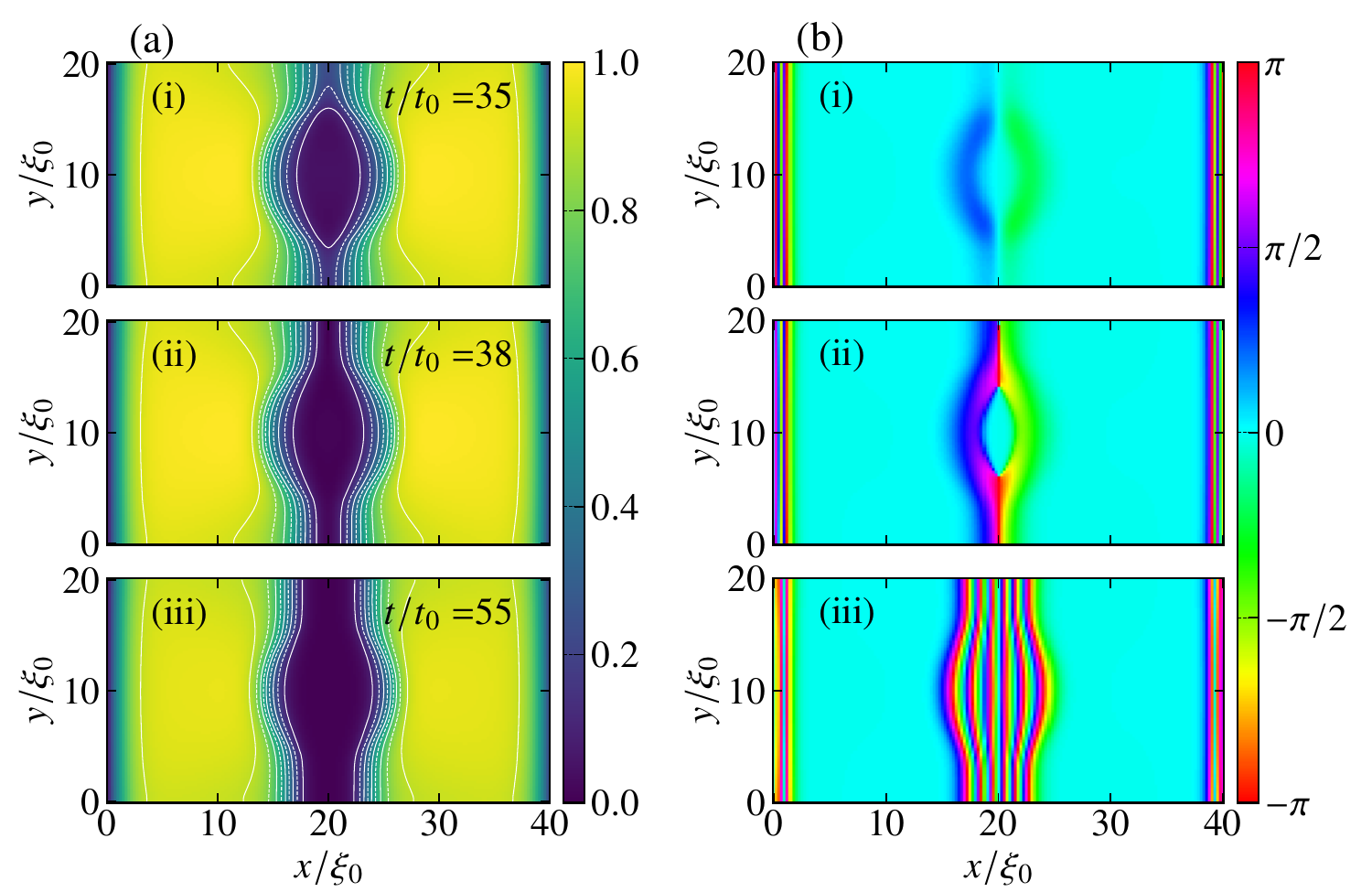}
\end{center}
\caption{
Dynamics of amplitude [(a) left panels] and phase [(b) right panels] of superconducting order parameter for $R_{\mathrm{init}}/\xi_0=6.0$, $j_{\mathrm{bias}}/j_{0}=0.28$, and $\kappa=10$. We take $t/t_{0}=35,38,55$ from top to bottom. In the left panels, the contour lines are plotted in the same manner as in Fig.~\ref{SSPDfig6}(a).
}
\label{SSPDfig7}
\end{figure}

Figure~\ref{SSPDfig7}(a) presents the dynamics of the amplitude of $\tilde{\Delta}(x,y)$ for $R_{\mathrm{init}}/\xi_0=6.0$, $j_{\mathrm{bias}}/j_{0}=0.28$, and $\kappa=10$. In this case, $R_{\mathrm{init}}$ is comparable to the penetration depth ($\lambda_0/\xi_0=10$). We find that the hotspot region simply expands and touches the upper and lower edges of the superconductor at the initial stage of the time evolution [see panel (i) of Fig.~\ref{SSPDfig7}(a) for $t/t_0=35$]. According to Fig.~\ref{SSPDfig2}, $\Delta V$ already takes nearly half of its maximum value at $t/t_0=35$. Small dark spots with size $\xi_0$ cannot be observed. When $R_{\mathrm{init}}$ is large, the current does not circulate around the hotspot. Then, the vortex--antivortex pairs are less likely to form.

The phase of $\tilde{\Delta}(x,y)$ is also shown in Fig.~\ref{SSPDfig7}(b). No $2\pi$ rotation of the phase occurs in the initial stage of the time evolution, although the normal region has almost been formed at $t/t_{0}=35$. At $t/t_0=38$ and $55$, we observe singularities in the phase. These should be interpreted with caution, since they are likely numerical artifacts arising from the fact that the order-parameter amplitude in the dark region has largely diminished, making it difficult to determine the phase accurately.

\subsection{Voltage change and dynamics of order parameter near the transition point between type-I and type-II superconducting states}

According to Ref.~\cite{Ota}, where the authors assumed $\kappa=2$, they concluded that the simple hotspot expansion is a possible mechanism of photon detection. To compare it with the present work, we examine the dependence of $\kappa$ on $\Delta V$ and dynamics of the superconducting order parameter. We take $\kappa=1$, which is close to the transition ($\kappa_{c}=1/\sqrt{2}$) from type-II to type-I superconductor. We expect that the formation of the vortex--antivortex pairs becomes unstable as $\kappa$ decreases. 

\begin{figure}[tb]
\begin{center}
\includegraphics[width=\hsize]{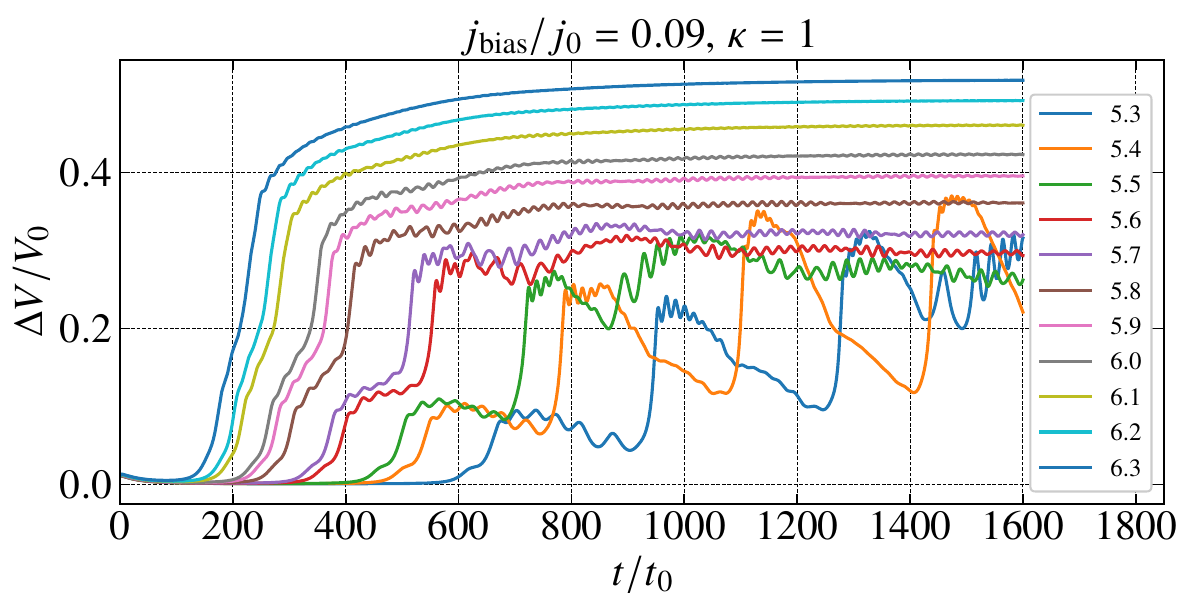}
\end{center}
\caption{
Time evolution of voltage change $\Delta V$. Various lines present the profiles with different values of $R_{\mathrm{init}}/\xi_0$. We take $j_{\mathrm{bias}}/j_{0}=0.09$ and $\kappa=1$.
}
\label{SSPDfig8}
\end{figure}

Figure~\ref{SSPDfig8} shows the time evolution of $\Delta V$ for $\kappa=1$. We take $j_{\mathrm{bias}}/j_{0}=0.09$, which is slightly lower than the critical current. We observe the oscillatory behavior on $\Delta V$ especially for small $R_{\mathrm{init}}$. However, the profile is quite different from the previous one in Fig.~\ref{SSPDfig2} for $\kappa=10$: faster and slower oscillations coexist.

\begin{figure}[tb]
\begin{center}
\includegraphics[width=\hsize]{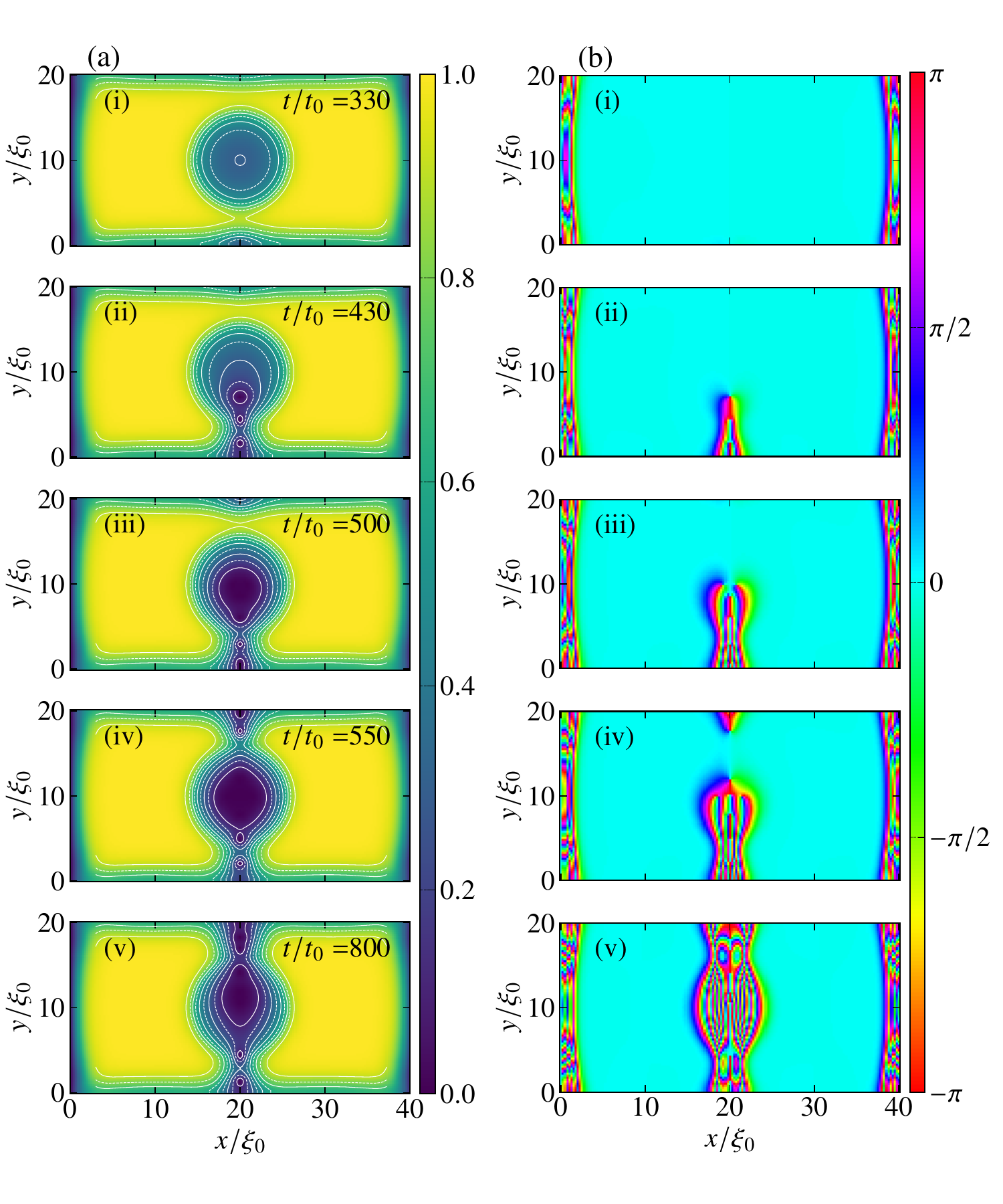}
\end{center}
\caption{
Dynamics of amplitude [(a) left panels] and phase [(b) right panels] of superconducting order parameter for $R_{\mathrm{init}}/\xi_0=5.6$, $j_{\mathrm{bias}}/j_{0}=0.09$, and $\kappa=1$. We take $t/t_{0}=330,430,500,550,800$ from top to bottom. In the left panels, the contour lines are plotted in the same manner as in Fig.~\ref{SSPDfig6}(a).
}
\label{SSPDfig9}
\end{figure}

To examine the difference, we show the dynamics of the order parameter for $R_{\mathrm{init}}/\xi_0=5.6$ in Fig.~\ref{SSPDfig9} (see red line in Fig.~\ref{SSPDfig8}). In this case, $\Delta V$ shows a step-like structure with oscillation around $400\le t/t_0\le 550$. We find the presence of penetrating antivortices from the lower edge of the sample in the initial stage of time evolution [see Figs.~\ref{SSPDfig9}(a)(ii) and \ref{SSPDfig9}(b)(ii) for $t/t_0=430$]. We also find the accumulation of penetrating antivortices in the initial hotspot area [see Figs.~\ref{SSPDfig9}(a)(iii) and \ref{SSPDfig9}(b)(iii) for $t/t_0=500$]. The accumulation process occurs, when $\Delta V$ shows the step-like structure. The vortex--antivortex pair is not formed inside the hotspot.
The oscillation of $\Delta V$ around $400\le t/t_0\le 550$ originates from penetration of antivortices at the lower edge. 
We have numerically confirmed that each penetration process occurs during the interval between the local minimum and local maximum of $\Delta V$. The next accumulation process starts by the penetration of vortices from the upper edge of the sample. The collision among these penetrated vortices and accumulated antivortices in the hotspot spikes the voltage.

The oscillation of $\Delta V$ is observed for both of $\kappa=10$ and $\kappa=1$, but the origins of the oscillation differ between these two cases.

\subsection{Effects of spatially non-uniform current density on voltage change and dynamics of order parameter: a perspective for SWSPD}

A key difficulty in photon detection using wide superconducting strips is the intrinsic dark count arising from the non-uniform distribution of the transport current. It is therefore important to understand how our results change when the current flow becomes more spatially non-uniform in wider strips. To this end, we examine how the sample width $L_y$ affects the dynamics of the superconducting order parameter.

\begin{figure}[tb]
\begin{center}
\includegraphics[width=0.9\hsize]{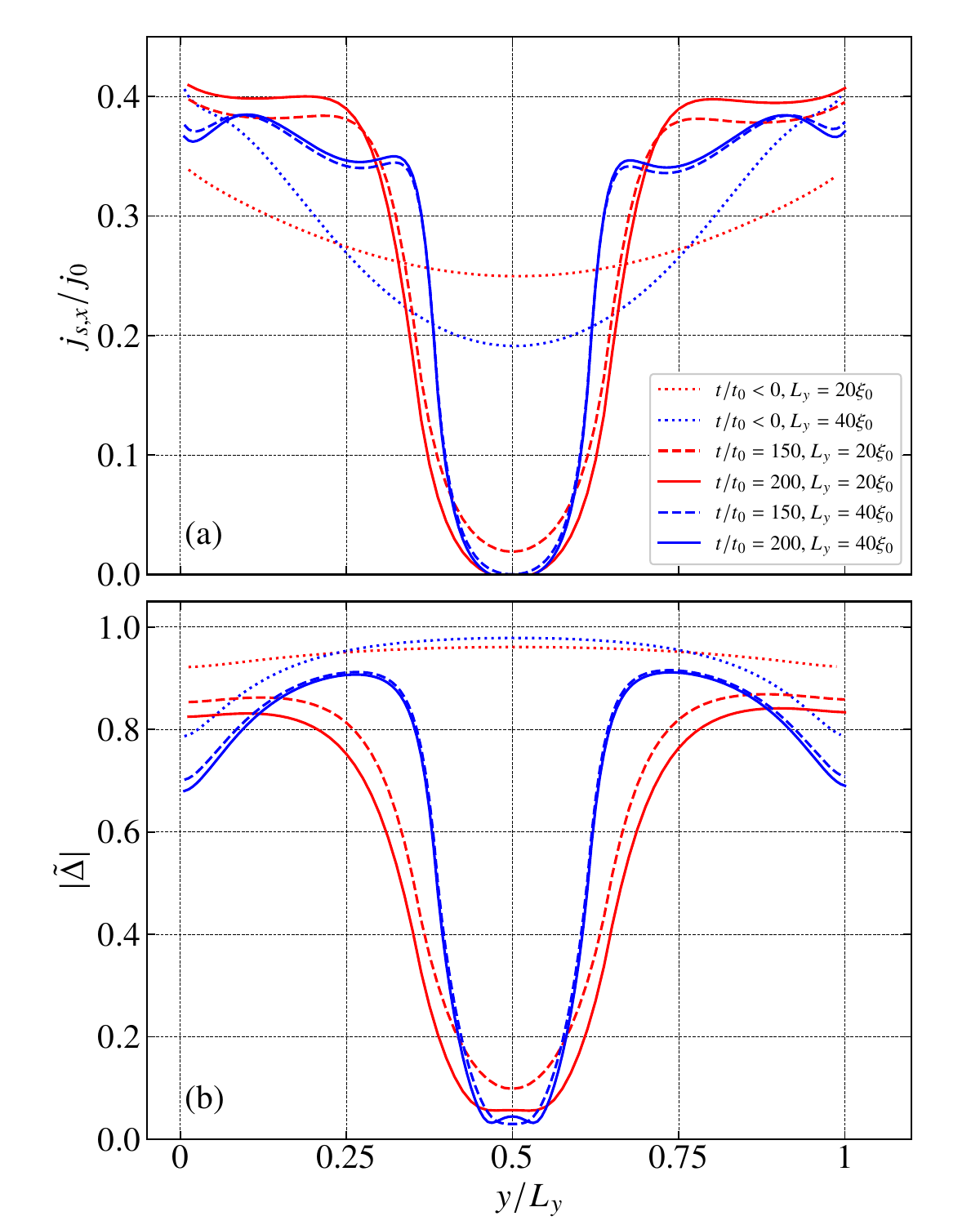}
\end{center}
\caption{
Spatial profile of the supercurrent density $j_{s,x}(x=L_{x}/2,y)$ [upper panel (a)] and the order-parameter amplitude $\left|\tilde{\Delta}(x=L_{x}/2,y)\right|$ [lower panel (b)] before photon injection (dotted lines) and after the formation of the initial hotspot (dashed and solid lines) for $L_{x}\times L_{y}=40\xi_0\times 20\xi_0$ (red lines) and $40\xi_0\times 40\xi_0$ (blue lines). Since the normal-current contribution $j_{n,x}$ is small under the present conditions, the integral of $j_{s,x}$ over $y$ is approximately equal to $j_{\mathrm{bias}}L_y$. The supercurrent density and the order parameter after the hotspot formation are calculated at $t/t_0=150$ (dashed lines) and $t/t_0=200$ (solid lines). The initial hotspot radius is taken as $R_{\mathrm{init}}/\xi_0 = 3.0$ ($4.7$) for $L_y = 20\xi_0$ ($40\xi_0$). We use $j_{\mathrm{bias}}/j_0=0.28$ and $\kappa=10$.
}
\label{SSPDfig10}
\end{figure}

In Fig.~\ref{SSPDfig10}(a), we compare the spatial profile of the supercurrent density, $j_{s,x}(L_x/2, y)$, before and after photon injection for a wider strip ($L_y = 40 \xi_0$) with those for $L_y = 20 \xi_0$. We note that all results presented so far were obtained for $L_y = 20\xi_0$. For the wider strip, we keep the same mesh size $\delta = 0.25\xi_0$, and in this geometry the critical current density lies between $j_c/j_0=0.285$ and $0.29$. For both strip widths, we set $j_{\mathrm{bias}}/j_0=0.28$ and $\kappa=10$.
In the situations shown here, the normal-current contribution is small
(although it is not explicitly plotted), so the integrated supercurrent
$\int j_{s,x}\,dy$ is approximately equal to $j_{\mathrm{bias}} L_y$. Before photon injection [dotted lines in Fig.~\ref{SSPDfig10}(a)], the screening effects---and hence the spatial non-uniformity of the current---become more pronounced as $L_y$ increases. For $L_y=20\xi_0$, the spatial profile of $j_{s,x}(L_x/2, y)$ remains close to $\cosh[(y-L_y/2)/\lambda]$ as expected from London theory. In contrast, for $L_y=40\xi_0$, an additional structure appears near the edges, reflecting the breakdown of the uniform-order-parameter assumption implicit in the London theory.
For a large strip width $L_y$, the edge currents expected from the London theory become substantially larger than those that can be sustained by the self-consistent superconducting state. 
In practice, however, such large edge currents suppress the order-parameter amplitude through nonlinear effects [see blue dotted line in Fig.~\ref{SSPDfig10}(b)], and this self-consistent suppression modifies the current distribution away from the $\cosh$ form near the edges.
As a result, a distinct edge structure appears for $L_y = 40\xi_0$.

After the photon injection [at $t/t_0 = 150$ and $t/t_0=200$, corresponding respectively to the dashed and solid lines in Fig.~\ref{SSPDfig10}(a)],  the supercurrent density inside the initial hotspot ($R_{\mathrm{init}}/\xi_0 = 3.0$ for $L_y = 20\xi_0$ and 4.7 for $L_y = 40\xi_0$) is reduced because the superconducting order parameter is suppressed by local heating, as shown in Fig.~\ref{SSPDfig10}(b).
In contrast, the current density surrounding the hotspot region increases, and is nearly saturated at $j_{s,x}(L_x/2,y)/j_0\lesssim 0.4$ for both strip widths. For $L_y = 40\xi_0$, the additional structures of the supercurrent density near the edges, which are already present before photon injection as mentioned in the previous paragraph, evolve into local minima, whose values decrease with time. This behavior contrasts with the edge supercurrent density for $L_y = 20\xi_0$. Such local minima correspond to the significant suppression of $\left|\tilde{\Delta}(L_x/2,y)\right|$ near the edges. We expect that the suppression affects the degree to which magnetic fluxes can penetrate.

\begin{figure}[tb]
\begin{center}
\includegraphics[width=\hsize]{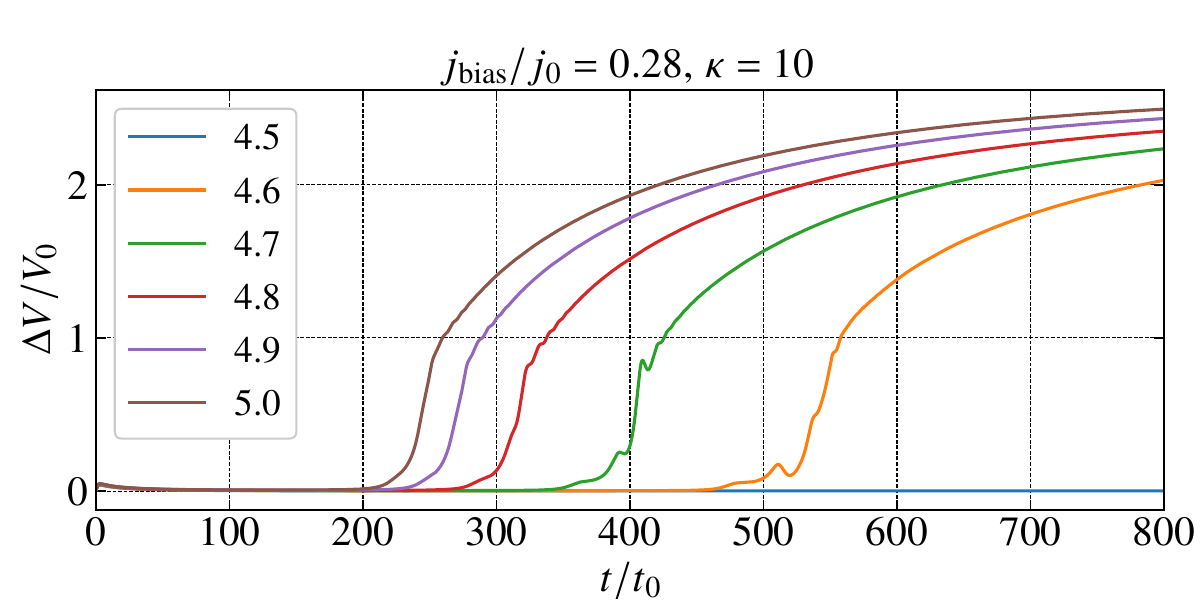}
\end{center}
\caption{
Time evolution of voltage change $\Delta V$ for $L_x\times L_y=40\xi_0\times 40\xi_0$. Various lines present the profiles with different values of $R_{\mathrm{init}}/\xi_0$. We take $j_{\mathrm{bias}}/j_0=0.28$ and $\kappa=10$.
}
\label{SSPDfig11}
\end{figure}

Based on these observations on the non-uniform current distribution for wider width cases, we examine the time-dependent voltage change $\Delta V$ and the dynamics of the superconducting order parameter for $L_{x}\times L_{y}=40\xi_0\times 40\xi_0$ to compare it with the previous results with $L_{x}\times L_{y}=40\xi_0\times 20\xi_0$. The voltage change is shown in Fig.~\ref{SSPDfig11}. The threshold for successful photon detection is located at $R_{\mathrm{init}}/\xi_0=4.6$. 
 The ratio between this threshold and the sample width $40\xi_0$ is $0.115$, which is smaller than the value $0.145$ for $L_{y}=20\xi_0$, with the corresponding threshold at $R_{\mathrm{init}}/\xi_0=2.9$. Thus, the increase in the threshold is weaker than linear, even when we linearly increase the sample width. Near the threshold value, the oscillation of $\Delta V$ is still observed, although the signature of the oscillation is less clear than in the case of narrower width. This oscillation disappears as we increase $R_{\mathrm{init}}$, and this behavior is the same as that in the narrower-width case. Our observation, that the dependence of the threshold on $L_y$ is weaker than linear and the vortex--antivortex pair formation is a possible mechanism of photon detection near the threshold, is very helpful for the realization of SWSPD.

\begin{figure}[tb]
\begin{center}
\includegraphics[width=\hsize]
{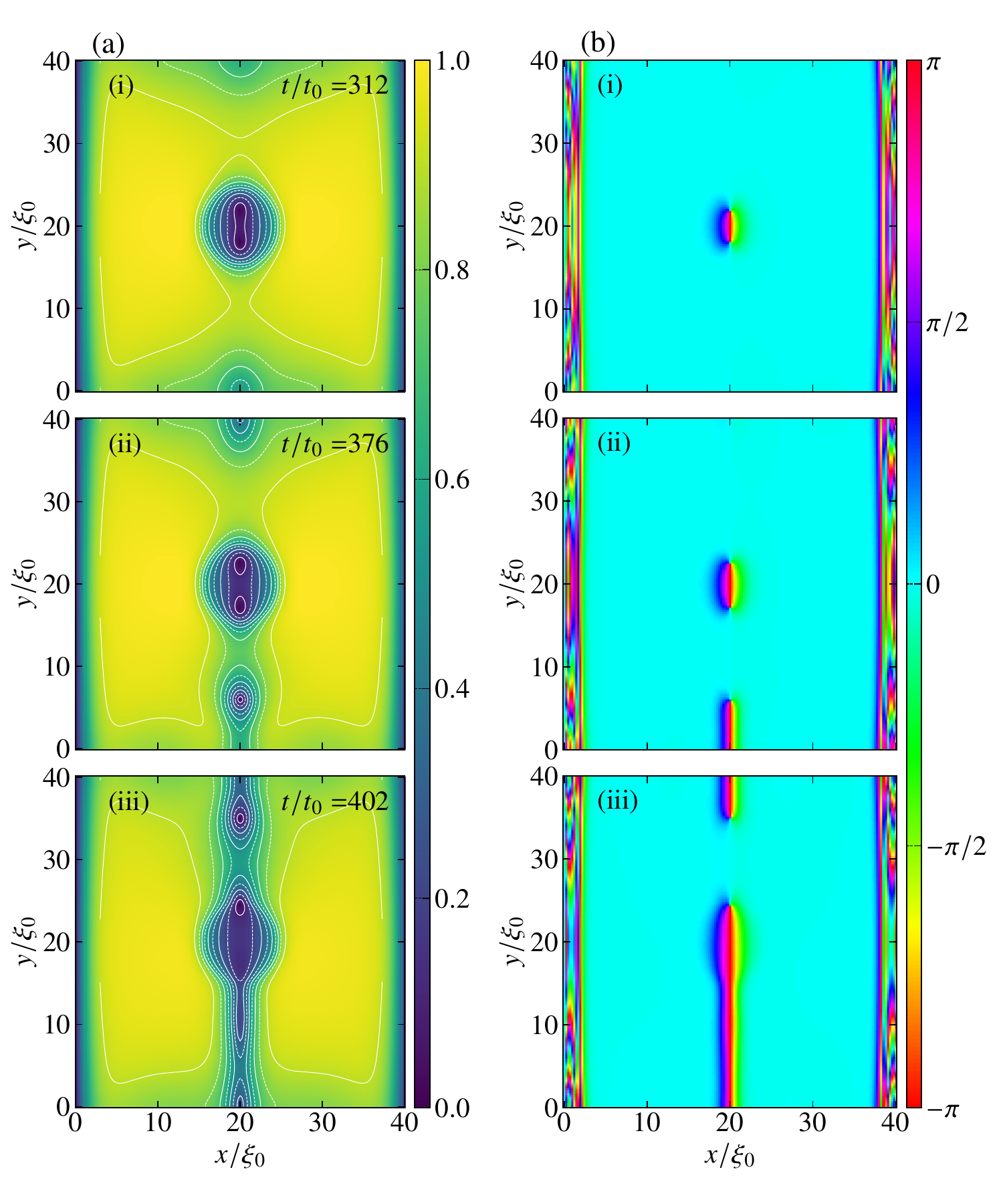}
\end{center}
\caption{
Dynamics of the amplitude and phase of the superconducting order parameter for $R_{\mathrm{init}}/\xi_0=4.7$ and $L_{x}\times L_{y}=40\xi_0\times 40\xi_0$. We take $t/t_{0}=312,376,402$ from top to bottom. In the left panels, the contour lines are plotted in the same manner as in Fig.~\ref{SSPDfig6}(a).
}
\label{SSPDfig12}
\end{figure}

The dynamics of the order parameter is presented in Fig.~\ref{SSPDfig12}. We find that a vortex--antivortex pair is also formed in the initial hotspot area [see Figs.~\ref{SSPDfig12}(a)(i) and \ref{SSPDfig12}(b)(i)], and then the antivortices penetrate the superconductor from the lower edge [see Figs.~\ref{SSPDfig12}(a)(ii) and \ref{SSPDfig12}(b)(ii)]. After that, vortices and antivortices penetrate from the upper and lower edges repeatedly [see Figs.~\ref{SSPDfig12}(a)(iii) and \ref{SSPDfig12}(b)(iii)]. The motion of the penetrating vortices and antivortices is more significant than that of the vortex--antivortex pair. As shown in Fig.~\ref{SSPDfig10}(b), $\left|\tilde{\Delta}(L_x/2,y)\right|$ is significantly suppressed near the edges for $L_y/\xi_0=40$, which faciliates the penetration of vortices and antivortices.

\subsection{Estimation of incident photon energy}

We finally comment on whether the results of this research have any predictive capability regarding the photon detection mechanism in a realistic parameter range. In the previous study~\cite{Ota}, the photon energy required for photon detection was two orders of magnitude larger than realistic values. Our analysis would overcome the previous difficulty. The incident photon energy is estimated by the following formula: $E_{\mathrm{photon}}=\pi R_{\mathrm{init}}^{2}dC_{v}\Delta T$, where $d$ is the sample thickness and $C_{v}$ is the heat capacity at the critical temperature. When we assume $\xi_0=5$ nm, $d=4$ nm, and $C_{v}=2.4$ mJ/cm$^{3}\cdot$K for $T_{c}=10$ K, which are suitable for NbN samples, we obtain a reasonable value $E_{\mathrm{photon}}=0.423$ eV for $R_{\mathrm{init}}/\xi_0=3.0$ and $\Delta T=T_{c}$. It is important for the fablication of SWSPD to know what value $R_{\mathrm{init}}$ converges to when the sample width increases. This is an interesting future work.

\section{Conclusion}
\label{conclusion}

We have performed the analysis of $\Delta V$ and $\tilde{\Delta}(x,y)$ on the basis of the TDGL and Maxwell equations to elucidate the photon detection mechanism of SSPD. For $\kappa=10$ and $R_{\mathrm{init}}$ larger than the threshold value, we have successfully observed the voltage change, leading to photon detection. Near the threshold value, we have observed the formation of a vortex--antivortex pair and the subsequent dynamics of the vortex and antivortex. The oscillatory behavior of $\Delta V$ characterizes the penetration of vortices and antivortices from outside of the superconductor. For large-$R_{\mathrm{init}}$ values where $\Delta V$ is smooth as a function of time, the formation mechanism of the normal region is hotspot-like. In the case of $\kappa$ close to the transition point from type-II to type-I superconductor, the oscillatory behavior of the voltage change is also observed, and this is also due to penetrating vortices (antivortices). Our results are qualitatively consistent with experimental results in the sense that a near-critical $R_{\mathrm{init}}$ value provides a reasonable value of $E_{\mathrm{photon}}$.
We hope that our analysis serves as a guideline for device fabrication.

Our numerical results and conclusions are somehow different from the previous works~\cite{Ota,Zotova}. Compared with the results in Ref.~\cite{Ota}, even for large-$\kappa$ values, we confirm the usual hotspot mechanism, if $R_{\mathrm{init}}$ is large enough. In Ref.~\cite{Ota} with a small $\kappa$ value, the injection of sufficiently large photon energy generates vortex--antivortex pairs, and this is different from our results. Our model settings and parameter selection are much more systematic. Compared with the results in Ref.~\cite{Zotova}, the ratio between $R_{\mathrm{init}}$ and $\xi_0$ for photon detection due to the presence of vortex--antivortex pairs is different. In Ref.~\cite{Zotova}, the authors treated the case in which $R_{\mathrm{init}}$ is comparable to $\xi_0$, and did not focus on the threshold of $R_{\mathrm{init}}$. On the other hand, there exists the threshold value of $R_{\mathrm{init}}$ in our case, and their comparable setting does not show the finite amount of voltage change.
In Ref.~\cite{Zotova}, the authors  focused on the threshold bias current above which the superconducting state becomes unstable, and thus did not pay particular attention to the threshold of $R_{\mathrm{init}}$.

In this paper, we treated the initial process of photon absorption as heating of a fixed region. We did not consider expansion of the initial hotspot region and subsequent heat relaxation of the hotspot into the heat sink, but we obtained a qualitatively reasonable result. We believe that these extra processes do not change our final results.

One of the open questions is about the origin of the response time. We have presented that our response time obtained for $R_{\mathrm{init}}/\xi_0=3.0$ is consistent with the experimental results. However, we do not yet understand what determines the magnitude of the response time. This is a future work.

\acknowledgements
We acknowledge Sanshiro Hatakeyama, Taro Yamashita, Shigehito Miki, Masahiro Yabuno, and Hirotaka Terai for fruitful discussion and providing us with useful information. We acknowledge financial support from KAKENHI projects No. JP23K17323 and No. JP24K17000. HM also acknowledges support from CSIS, Tohoku University.

\bibliography{sspd2}

@article{Bartolf,
  title = {Current-assisted thermally activated flux liberation in ultrathin nanopatterned NbN superconducting meander structures},
  author = {Bartolf, H. and Engel, A. and Schilling, A. and Il'in, K. and Siegel, M. and H{\"u}bers, H.-W. and Semenov, A.},
  year = 2010,
  journal = {Phys. Rev. B},
  volume = {81},
  pages = {024502},
  doi = {10.1103/PhysRevB.81.024502},
}

@article{Bulaevskii,
  title = {Vortex-assisted photon counts and their magnetic field dependence in single-photon superconducting detectors},
  author = {Bulaevskii, L. N. and Graf, Matthias J. and Kogan, V. G.},
  year = 2012,
  journal = {Phys. Rev. B},
  volume = {85},
  number = {1},
  pages = {014505},
  doi = {10.1103/PhysRevB.85.014505},
  url = {https://link.aps.org/doi/10.1103/PhysRevB.85.014505}
}

@article{Bulaevskii2,
  title = {Vortex-induced dissipation in narrow current-biased thin-film superconducting strips},
  author = {Bulaevskii, L. N. and Graf, Matthias J. and Batista, C. D. and Kogan, V. G.},
  year = 2011,
  journal = {Phys. Rev. B},
  volume = {83},
  pages = {144526},
  doi = {10.1103/PhysRevB.83.144526},
}

@article{Engel,
  title = {Numerical analysis of detection-mechanism models of superconducting nanowire single-photon detector},
  author = {Engel, Andreas and Schilling, Andreas},
  year = 2013,
  journal = {J. Appl. Phys.},
  volume = {114},
  number = {21},
  pages = {214501},
  doi = {10.1063/1.4836878},
  url = {https://pubs.aip.org/jap/article/114/21/214501/309038/Numerical-analysis-of-detection-mechanism-models},
  urldate = {2025-11-03},
  langid = {english}
}

@article{Engel2,
  title = {Detection mechanism of superconducting nanowire single-photon detectors},
  author = {Engel, Andreas and Renema, J. J. and Semenov, A},
  year = 2015,
  journal = {Supercond. Sci. Teconol.},
  volume = {28},
  pages = {114003},
  doi = {10.1088/0953-2048/28/11/114003},
  langid = {english}
}

@article{Ge,
  title = {Controlled {{Generation}} of {{Quantized Vortex}}--{{Antivortex Pairs}} in a {{Superconducting Condensate}}},
  author = {Ge, Jun-Yi and Gladilin, Vladimir N. and Tempere, Jacques and Devreese, Jozef and Moshchalkov, Victor V.},
  year = 2017,
  journal = {Nano Lett.},
  volume = {17},
  number = {8},
  pages = {5003--5007},
  publisher = {American Chemical Society},
  doi = {10.1021/acs.nanolett.7b02180},
  url = {https://doi.org/10.1021/acs.nanolett.7b02180},
  urldate = {2023-03-08}
}

@article{Goltsman,
  title = {{Picosecond superconducting single-photon optical detector}},
  author = {Gol'tsman, G. N. and Okunev, O. and Chulkova, G. and Lipatov, A. and Semenov, A. and Smirnov, K. and Voronov, B. and Dzardanov, A. and Williams, C. and Sobolewski, Roman},
  year = 2001,
  journal = {Appl. Phys. Lett.},
  volume = {79},
  number = {6},
  pages = {705--707},
  doi = {10.1063/1.1388868},
  url = {https://pubs.aip.org/apl/article/79/6/705/515874/Picosecond-superconducting-single-photon-optical},
  urldate = {2025-11-03},
  langid = {english}
}

@article{Goltsman2,
  title = {Ultrafast superconducting single-photon detectors for near-infrared-wavelength quantum communications},
  author = {Gol'tsman, G. N. and Korneev, A. and Rubtsova, I. and Milostnaya, I. and Chulkova, G. and Minaeva, O. and Smirnov, K. and Voronov, B. and S{\l{}}ysz, W. and Pearlman, A. and Verevkin, A. and Sobolewski, Roman},
  year = 2005,
  journal = {phys. stat. sol. (c)},
  volume = {2},
  number = {5},
  pages = {1480--1488},
  doi = {10.1002/pssc.200460829},
  url = {https://onlinelibrary.wiley.com/doi/10.1002/pssc.200460829},
  urldate = {2025-11-03},
  copyright = {http://onlinelibrary.wiley.com/termsAndConditions\#vor},
  langid = {english}
}

@article{Hadfield,
  title = {Single-photon detectors for optical quantum information applications},
  author = {Hadfield, Robert H.},
  year = 2009,
  journal = {Nat. Photonics},
  volume = {3},
  number = {12},
  pages = {696--705},
  doi = {10.1038/nphoton.2009.230},
  url = {https://www.nature.com/articles/nphoton.2009.230},
  urldate = {2025-11-03},
  copyright = {http://www.springer.com/tdm},
  langid = {english}
}

@article{He,
  author       = {Daien He and Leif Bauer and Sathwik Bharadwaj and Zubin Jacob},
  title        = {{Unified Theory of Dark Count Rate and System Detection Efficiency for NbN and WSi-Based Superconducting Single-Photon Detectors}},
  journal      = {arXiv:2508.10816 [cond-mat.supr-con]},
  year         = {2025},
  url          = {https://arxiv.org/abs/2508.10816}
}

@article{Hofherr,
  title = {Intrinsic detection efficiency of superconducting nanowire single-photon detectors with different thicknesses},
  author = {Hofherr, M. and Rall, D. and Ilin, K. and Siegel, M. and Semenov, A. and H{\"u}bers, H.-W. and Gippius, N. A.},
  year = 2010,
  journal = {J. Appl. Phys.},
  volume = {108},
  number = {1},
  pages = {014507},
  doi = {10.1063/1.3437043},
  url = {https://doi.org/10.1063/1.3437043},
  urldate = {2025-10-28}
}

@article{Kadin,
  title = {Photon-assisted vortex depairing in two-dimensional superconductors},
  author = {Kadin, A. M. and Leung, M. and Smith, A. D.},
  year = 1990,
  journal = {Phys. Rev. Lett.},
  volume = {65},
  number = {25},
  pages = {3193--3196},
  publisher = {American Physical Society},
  doi = {10.1103/PhysRevLett.65.3193},
  url = {https://link.aps.org/doi/10.1103/PhysRevLett.65.3193},
  urldate = {2023-03-08}
}

@article{Kadin2,
  title = {Photon-stimulated production of vortex-antivortex pairs in thin superconducting films},
  author = {Kadin, A. M. and Leung, M. and Smith, A. D. and Murduck, J. M.},
  year = 1991,
  journal = {Phys. B Condens. Matter},
  volume = {169},
  number = {1},
  pages = {681--682},
  doi = {10.1016/0921-4526(91)90386-S},
  url = {https://www.sciencedirect.com/science/article/pii/092145269190386S},
  urldate = {2023-03-08},
  langid = {english}
}

@article{Kadin3,
  title = {Nonequilibrium photon-induced hotspot: {{A}} new mechanism for photodetection in ultrathin metallic films},
  shorttitle = {Nonequilibrium photon-induced hotspot},
  author = {Kadin, A. M. and Johnson, M. W.},
  year = 1996,
  journal = {Appl. Phys. Lett.},
  volume = {69},
  number = {25},
  pages = {3938--3940},
  doi = {10.1063/1.117576},
  url = {https://pubs.aip.org/apl/article/69/25/3938/518885/Nonequilibrium-photon-induced-hotspot-A-new},
  urldate = {2025-11-03},
  langid = {english}
}

@article{Kato,
  title = {Effects of the surface boundary on the magnetization process in type-{{II}} superconductors},
  author = {Kato, Ryuzo and Enomoto, Yoshihisa and Maekawa, Sadamichi},
  year = 1993,
  journal = {Phys. Rev. B},
  volume = {47},
  number = {13},
  pages = {8016--8024},
  doi = {10.1103/PhysRevB.47.8016},
  url = {https://link.aps.org/doi/10.1103/PhysRevB.47.8016},
  urldate = {2023-01-18},
  langid = {english}
}

@article{Korneeva,
  title = {Optical {{Single-Photon Detection}} in {{Micrometer-Scale NbN Bridges}}},
  author = {Korneeva, {\relax Yu}. P. and Vodolazov, D. {\relax Yu}. and Semenov, A. V. and Florya, I. N. and Simonov, N. and Baeva, E. and Korneev, A. A. and Goltsman, G. N. and Klapwijk, T. M.},
  year = 2018,
  journal = {Phys. Rev. Appl.},
  volume = {9},
  number = {6},
  pages = {064037},
  doi = {10.1103/PhysRevApplied.9.064037},
  url = {https://link.aps.org/doi/10.1103/PhysRevApplied.9.064037},
  urldate = {2025-11-03},
  langid = {english}
}

@article{Kozorezov,
  title = {Quasiparticle recombination in hotspots in superconducting current-carrying nanowires},
  author = {Kozorezov, A. G. and Lambert, C. and Marsili, F. and Stevens, M. J. and Verma, V. B. and Stern, J. A. and Horansky, R. and Dyer, S. and Duff, S. and Pappas, D. P. and Lita, A. and Shaw, M. D. and Mirin, R. P. and Nam, Sae Woo},
  year = 2015,
  journal = {Phys. Rev. B},
  volume = {92},
  pages = {064504--064504},
  doi = {10.1103/PhysRevB.92.064504},
  langid = {english},
}

@article{Marsili,
  title = {Hotspot rexation dynamics in a current-carrying superconductor},
  author = {Marsili, F. and Stevens, M. J. and Kozorezov, A. G. and Verma, V. B. and Lambert, Colin and Stern, J. A. and Horansky, R. D. and Dyer, S. and Duff, S. and Pappas, D. P. and Lita, A. E. and Shaw, M. D. and Mirin, R. P. and Nam, Sae Woo},
  year = 2016,
  journal = {Phys. Rev. B},
  volume = {93},
  pages = {094518--094518},
  doi = {10.1103/PhysRevB.93.094518},
  langid = {english},
}

@article{Miki,
  title = {Compactly packaged superconducting nanowire single-photon detector with an optical cavity for multichannel system},
  author = {Miki, Shigehito and Takeda, Masanori and Fujiwara, Mikio and Sasaki, Masahide and Wang, Zhen},
  year = 2009,
  journal = {Opt. Express},
  volume = {17},
  number = {26},
  pages = {23557--23564},
  publisher = {Optica Publishing Group},
  doi = {10.1364/OE.17.023557},
  url = {https://opg.optica.org/oe/abstract.cfm?uri=oe-17-26-23557},
  urldate = {2025-11-03},
  copyright = {\copyright{} 2009 OSA},
  langid = {english}
}

@article{Miki2,
  title = {Multichannel {{SNSPD}} system with high detection efficiency at telecommunication wavelength},
  author = {Miki, Shigehito and Yamashita, Taro and Fujiwara, Mikio and Sasaki, Masahide and Wang, Zhen},
  year = 2010,
  journal = {Opt. Lett.},
  volume = {35},
  number = {13},
  pages = {2133--2135},
  publisher = {Optica Publishing Group},
  doi = {10.1364/OL.35.002133},
  url = {https://opg.optica.org/ol/abstract.cfm?uri=ol-35-13-2133},
  urldate = {2025-11-03},
  copyright = {\copyright{} 2010 Optical Society of America},
  langid = {english}
}

@article{Miki3,
  title = {High performance fiber-coupled {{NbTiN}} superconducting nanowire single photon detectors with {{Gifford-McMahon}} cryocooler},
  author = {Miki, Shigehito and Yamashita, Taro and Terai, Hirotaka and Wang, Zhen},
  year = 2013,
  journal = {Opt. Express},
  volume = {21},
  number = {8},
  pages = {10208--10214},
  publisher = {Optica Publishing Group},
  doi = {10.1364/OE.21.010208},
  url = {https://opg.optica.org/oe/abstract.cfm?uri=oe-21-8-10208},
  urldate = {2025-11-03},
  copyright = {\copyright{} 2013 OSA},
  langid = {english}
}

@article{Ota,
  title = {Full {{Numerical Simulations}} of {{Dynamical Response}} in {{Superconducting Single-Photon Detectors}}},
  author = {Ota, Y. and Kobayashi, K. and Machida, M. and Koyama, T. and Nori, F.},
  year = 2013,
  journal = {IEEE Trans. Appl. Supercond.},
  volume = {23},
  number = {3},
  pages = {2201105--2201105},
  doi = {10.1109/TASC.2013.2248871},
  url = {http://ieeexplore.ieee.org/document/6470656/},
  urldate = {2025-11-03},
  copyright = {https://ieeexplore.ieee.org/Xplorehelp/downloads/license-information/IEEE.html}
}

@article{Qiu,
  title = {Numerical study of the phase slip in two-dimensional superconducting strips},
  author = {Qiu, Chunyin and Qian, Tuezheng},
  year = 2008,
  journal = {Phys. Rev. B},
  volume = {177},
  pages = {174517},
  doi = {10.1103/PhysRevB.77.174517},
  langid = {english}
}

@article{Saman,
  title = {Probabilistic vortex crossing criterion for superconducting nanowire single-photon detectors},
  author = {Jahani, Saman and Yang, Li-Ping and Buganza Tepole, Adri{\'a}n and Bardin, Joseph C. and Tang, Hong X. and Jacob, Zubin},
  year = 2020,
  journal = {J. Appl. Phys.},
  volume = {127},
  pages = {143101},
  doi = {10.1063/1.5132961},
  langid = {english}
}

@article{Semenov,
  title = {Quantum detection by current carrying superconducting film},
  author = {Semenov, Alex D. and Gol'tsman, Gregory N. and Korneev, Alexander A.},
  year = 2001,
  journal = {Phys. C Supercond.},
  volume = {351},
  number = {4},
  pages = {349--356},
  doi = {10.1016/S0921-4534(00)01637-3},
  url = {https://linkinghub.elsevier.com/retrieve/pii/S0921453400016373},
  urldate = {2025-11-03},
  langid = {english}
}

@article{Semenov2,
  title = {Spectral cut-off in the efficiency of the resistive state formation caused by absorption of a single-photon in current-carrying superconducting nano-strips},
  author = {Semenov, A. and Engel, A. and H{\"u}bers, H.-W. and Il'in, K. and Siegel, M.},
  year = 2005,
  journal = {Euro. Phys. J. B},
  volume = {47},
  pages = {495--501},
  doi = {10.1140/epjb/e2005-00351-8},
  langid = {english}
}

@article{Vodolazov,
  title = {Single-{{Photon Detection}} by a {{Dirty Current-Carrying Superconducting Strip Based}} on the {{Kinetic-Equation Approach}}},
  author = {Vodolazov, D. {\relax Yu}.},
  year = 2017,
  journal = {Phys. Rev. Appl.},
  volume = {7},
  number = {3},
  pages = {034014},
  doi = {10.1103/PhysRevApplied.7.034014},
  url = {https://link.aps.org/doi/10.1103/PhysRevApplied.7.034014},
  urldate = {2025-11-03},
  copyright = {http://link.aps.org/licenses/aps-default-license},
  langid = {english}
}

@article{Vodolazov2,
  title = {Vortex-assisted mechanism of photon counting in a superconducting nanowire single-photon detector revealed by external magnetic field},
  author = {Vodolazov, D. {\relax Yu}. and Korneeva, {\relax Yu}. P. and Semenov, A. V. and Korneev, A. A. and Goltsman, G. N.},
  year = 2015,
  journal = {Phys. Rev. B},
  volume = {92},
  pages = {104503},
  doi = {10.1103/PhysRevB.92.104503},
  langid = {english}
}

@article{Yabuno,
  title = {Superconducting wide strip photon detector with high critical current bank structure},
  author = {Yabuno, Masahiro and China, Fumihiro and Terai, Hirotaka and Miki, Shigehito},
  year = 2023,
  journal = {Opt. Quantum},
  volume = {1},
  number = {1},
  pages = {26--34},
  publisher = {Optica Publishing Group},
  doi = {10.1364/OPTICAQ.497675},
  url = {https://opg.optica.org/opticaq/abstract.cfm?uri=opticaq-1-1-26},
  urldate = {2025-11-03},
  copyright = {\copyright{} 2023 Optica Publishing Group},
  langid = {english}
}

@article{Yamashita2,
  title = {Temperature {{Dependent Performances}} of {{Superconducting Nanowire Single-Photon Detectors}} in an {{Ultralow-Temperature Region}}},
  author = {Yamashita, Taro and Miki, Shigehito and Qiu, Wei and Fujiwara, Mikio and Sasaki, Masahide and Wang, Zhen},
  year = 2010,
  journal = {Appl. Phys. Express},
  volume = {3},
  number = {10},
  pages = {102502},
  doi = {10.1143/APEX.3.102502},
  url = {https://iopscience.iop.org/article/10.1143/APEX.3.102502},
  urldate = {2025-11-03},
  langid = {english}
}

@article{Yamashita4,
  title = {Origin of intrinsic dark count in superconducting nanowire single-photon detectors},
  author = {Yamashita, T. and Miki, S. and Makise, K. and Qiu, W. and Terai, H. and Fujiwara, M. and Sasaki, M. and Wang, Z.},
  year = 2011,
  journal = {Appl. Phys. Lett.},
  volume = {99},
  number = {16},
  pages = {161105},
  doi = {10.1063/1.3652908},
  url = {http://aip.scitation.org/doi/10.1063/1.3652908}
}

@article{Yamashita3,
  title = {Performances of {{Fiber-Coupled Superconducting Nanowire Single-Photon Detectors Measured}} at {{Ultralow Temperature}}},
  author = {Yamashita, T. and Miki, S. and Qiu, W. and Fujiwara, M. and Sasaki, M. and Wang, Z.},
  year = 2011,
  journal = {IEEE Trans. Appl. Supercond.},
  volume = {21},
  number = {3},
  pages = {336--339},
  doi = {10.1109/TASC.2010.2091244},
  url = {http://ieeexplore.ieee.org/document/5643197/},
  urldate = {2025-11-03},
  copyright = {https://ieeexplore.ieee.org/Xplorehelp/downloads/license-information/IEEE.html}
}

@article{Yamashita5,
  title = {Low-filling-factor superconducting single photon detector with high system detection efficiency},
  author = {Yamashita, Taro and Miki, Shigehito and Terai, Hirotaka and {Zhen Wang}},
  year = 2013,
  journal = {Opt. Express},
  volume = {21},
  number = {22},
  pages = {27177--27184},
  publisher = {Optica Publishing Group},
  doi = {10.1364/OE.21.027177},
  url = {https://opg.optica.org/oe/abstract.cfm?uri=oe-21-22-27177},
  urldate = {2025-11-03},
  copyright = {\copyright{} 2013 Optical Society of America},
  langid = {english}
}

@article{Yamashita,
  title = {Recent {{Progress}} and {{Application}} of {{Superconducting Nanowire Single-Photon Detectors}}},
  author = {Yamashita, Taro and Miki, Shigehito and Terai, Hirotaka},
  year = 2017,
  journal = {IEICE Trans. Electron.},
  volume = {E100.C},
  number = {3},
  pages = {274--282},
  doi = {10.1587/transele.E100.C.274},
  url = {https://www.jstage.jst.go.jp/article/transele/E100.C/3/E100.C\_274/\_article},
  urldate = {2025-11-03},
  langid = {english}
}

@article{Zotova,
  title = {Photon detection by current-carrying superconducting film: {{A}} time-dependent {{Ginzburg-Landau}} approach},
  author = {Zotova, A. N. and Vodolazov, D. {\relax Yu}.},
  year = 2012,
  journal = {Phys. Rev. B},
  volume = {85},
  number = {2},
  pages = {024509},
  doi = {10.1103/PhysRevB.85.024509},
  url = {https://link.aps.org/doi/10.1103/PhysRevB.85.024509}
}

@article{Zotova2,
  title = {Intrinsic detection efficiency of superconducting nanowire single photon detector in the modified hot spot model},
  author = {Zotova, A. N. and Vodolazov, D. {\relax Yu}.},
  year = 2014,
  journal = {Supercond. Sci. Technol.},
  volume = {27},
  pages = {125001},
  doi = {10.1088/0953-2048/27/12/125001},
}

\end{document}